\documentclass[12pt]{article}
\usepackage[dvips]{color}
\usepackage{epsfig}
\usepackage{amsmath}
\usepackage{graphicx}

\begin{document}

\title{\bf Effects of dark energy on $P-V$ criticality and efficiency of charged Rotational black hole}

\author{{Kh. Jafarzade $^{a}$ \thanks{Email: kh.Jafarzadeh@stu.umz.ac.ir} \hspace{1mm},
J. Sadeghi $^{a}$\thanks{Email:
pouriya@ipm.ir}} \\
$^a${\small {\em Sciences Faculty, Department of Physics, University
of Mazandaran,}}\\{\small {\em P. O. Box 47415-416, Babolsar,
Iran}}} \maketitle
\begin{abstract}
In this paper, we study $P-V$ criticality of Kerr-Newman $AdS$ black
hole with a quintessence field. We calculate critical quantities and
show that for the equation state parameter $\omega= -\frac{1}{3}$,
the obtained universal ratio ($\frac{P_{c}\upsilon_{c}}{T_{c}}$) is
quite same as Kerr-Newman $AdS$ black hole without dark energy
parameter. We investigate the influence of quintessence field
$\alpha$, equation state parameter $\omega$ and angular momentum $J$
on the efficiency $\eta$. We find that $\eta$ is increased by
increasing $J$ and $\alpha$ and decreasing charge $Q$ of black hole.
We show when $\omega$ increases from $-1$ to $-\frac{1}{3}$ the
efficiency decreases. Also we study ratio $\frac{\eta}{\eta_{C}}$
(which $\eta_{C}$ is the Carnot efficiency) and see that the second
law of the thermodynamics is satisfied by special values of $J$ and
$\alpha$ and holds for any value of $Q$. We notice that in this case
by increasing $\omega$ from $-1$ to
$-\frac{1}{3}$ the range of $J$ and $\alpha$ increases.
\end{abstract}

\section{Introduction}
As we know black holes play an important role in physics especially
quantum gravity. One of the interesting methods  to study the
quantum gravity is black hole thermodynamics in AdS spacetime [1,2].
First time Hawking and Bekenstein investigated  the black hole
thermodynamic. They found that all laws of black hole mechanics are
similar to ordinary thermodynamics where surface gravity,  mass  and
area of black hole are related to the temperature, energy and
entropy respectively [3]. The study  of black hole thermodynamic
will be interesting  with presence of a negative cosmological
constant. Because such a cosmological constant play important role
in holography and $AdS/CFT$. From AdS/CFT point of view,
asymptotically AdS black hole spacetimes admit a gauge duality
description with dual thermal field theory. Such theory lead us to
interesting phenomenon which is called phase transition. The phase
transition plays a key role to study thermodynamical properties of a
system near critical point [4-8]. For the first time, the phase
transition was studied by Hawking and Page [9]. They found a phase
transition between Schwarzschild-AdS black holes and thermal
radiation which is known as Hawking-Page phase transition. After
that, a lot of studies done in context of AdS black hole
thermodynamics especially  in extended phase space [10-14]. In AdS
black hole thermodynamics, the cosmological constant of the
spacetime treat as pressure and its conjugate quantity as a
thermodynamic volume. But in extended phase space, it appear as a
thermodynamical variable in the first law of black hole
thermodynamics. Here, remarkable matter is that in extended phase
space the mass of black hole $M$ is not related to energy but it is
associated with enthalpy by $M=H\equiv U+PV$ [15]. So the first Law
of thermodynamics is expressed by,
\begin{equation}
dM=TdS+VdP+\Phi dQ+\Omega dJ.
\end{equation}
The thermodynamical properties will be very interesting with
presence a dark energy parameter. Recently observations show the
accelerating expansion of the universe which may be due to dark
energy [16,17]. The dark energy imply a state with the negative
pressure. There are two proposed forms for dark energy.  The first
candidate which is the simplest explanation for dark energy is
cosmological constant and the second  is called quintessence which
is described by the state equation $\omega=\frac{P}{\rho}$ where $P$
and $ \rho $ are pressure and energy density respectively. The state
parameter $\omega$ be restricted $-1<\omega<1$ which the case of
$\omega = \frac{1}{3}$ represents radiation and case of $\omega = 0$
represents dust around black hole. The range of $
-1<\omega<-\frac{1}{3} $ causes the acceleration and $ \omega =-1$
covers the cosmological constant term [18-23]. As we know [24] dark
energy forms $70$ percent of the universe and black holes are also
part of this universe. This fact makes a motivation for studying
black holes surrounded by dark energy. Such black holes play the
crucial role in cosmology and it is very interesting to know
influence of dark energy on thermodynamical properties of black
holes. The first time in 2003, Kiselev obtained the Einstein
equation's solution for the quintessence matter around schwarzschild
black hole. This solution is written in terms of $\omega$ and
$\alpha$ where $\alpha$ is a normalization factor that represents
the intensity of the quintessence field related to the black hole
[17]. The Kiselev black hole for different situations was studied in
Ref. [25-34]. The rotation Kiselev and Kerr-Newman kiselev solution
was determined in Ref. [35-38] and also Kerr- Newman-AdS solution in
the quintessence was obtained in [39]. Then some efforts have been
made in context of thermodynamic and phase transition for Kiselev
and Kerr Kiselev black hole [18,39].\\
After studying of phase transition, it is interesting to define
classical cycles for black holes like usual thermodynamic systems.
It means that when small black hole translate to large black hole we
need again the large black hole translate to small black hole, in
other words the system must be back to primary state. For the first
time the holographic heat engine has been studied for charge $AdS$
black hole [40]. Thereafter some effort has been made for different
situations [41-56]. Authors investigated the holographic heat engine
for charge $AdS$ black hole with presence dark energy parameter in
Ref. [23]. Also rotational black hole as heat engine was studied in
Ref. [57-58]. In this paper, we are going to study holographic heat
engine for charged rotational black hole with presence dark energy
parameter and discuss about the influence of quintessence field on
the $P-V$ criticality and efficiency of a such black hole.\\
The outline of the paper is as follows. In section II, we
investigate $P-V$ criticality of Kerr-Newman-AdS black hole solution
with present dark energy parameter. In section III, we consider
corresponding black hole as heat engine and study the influence of
quintessence field on the efficiency. Finally, in section IV we have
some results and conclusion.
\section{P-V criticality of Kerr-Newman-AdS black hole with quintessence field}
The Kerr-Newman AdS metric in quintessence matter is expressed as
follows [28],
\begin{equation}
ds^{2}=\frac{\Sigma^{2}}{\Delta_{r}}dr^{2}+\frac{\Sigma^{2}}{\Delta_{\theta}}d\theta^{2}+\frac{\Delta_{\theta}sin^{2}\theta}{\Sigma^{2}}(a\frac{dt}{\Xi}-(r^{2}-a^{2})\frac{d\phi}{\Xi})^{2}-\frac{\Delta_{r}}{\Sigma^{2}}(\frac{dt}{\Xi}-a\sin^{2}\theta
\frac{d\phi}{\Xi})^{2},
\end{equation}

where
\begin{eqnarray}
\Delta_{r}=r^{2}-2Mr+a^{2}+Q^{2}+\frac{r^{2}}{\ell^{2}}(r^{2}+a^{2})-\alpha
r^{1-3\omega},\nonumber
\\&& \hspace{-98mm}
\Delta_{\theta}=1-\frac{a^{2}}{\ell^{2}}\cos^{2}\theta~~~~~~~~
\Xi=1-\frac{a^{2}}{\ell^{2}}
~~~~~~~~\Sigma^{2}=r^{2}+a^{2}\cos^{2}\theta,
\end{eqnarray}
 where  mass $M$, charge $Q$, the angular momentum $J$ are
related to parameters $m$, $q$, and $a$ respectively by,
\begin{equation}
M=\frac{m}{\Xi^{2}},~~~~Q=\frac{q}{\Xi},~~~~~~J=\frac{am}{\Xi^{2}}
\end{equation}
as pervious mentioned  $\alpha$ and $\omega$ are normalization
factor and state parameter respectively. The cosmological constant
$\Lambda=-\frac{3}{\ell^{2}}$ interpret as a thermodynamic pressure
$P$ which is given by,
\begin{equation}
P=-\frac{\Lambda}{8\pi}=\frac{3}{8\pi \ell^{2}}.
\end{equation}
The mass of black hole is determined by $\Delta_{r}(r_{+})=0$, so
\begin{equation}
M=\frac{(r_{+}^{2}+a^{2})(r_{+}^{2}+\ell^{2})+Q^{2}\ell^{2}-\alpha
\ell^{2}r^{1-3\omega}}{2r_{+}\ell^{2}},
\end{equation}
the area of event horizon is given by,
\begin{equation}
A=4\pi\frac{(r_{+}^{2}+a^{2})}{\Xi},
\end{equation}
so The Bakenstein-Hawking entropy is expressed by
\begin{equation}
S=\frac{A}{4}=\pi\frac{(r_{+}^{2}+a^{2})}{\Xi}.
\end{equation}
The Hawking temperature which is related to surface gravity is
defined by,
\begin{equation}
T_{H}=\frac{\kappa(r_{+})}{2\pi}=\lim_{\theta=0, r\rightarrow
r_{+}}\frac{\partial_{r}\sqrt{g_{tt}}}{2\pi\sqrt{g_{rr}}},
\end{equation}
thus, the thermal temperature of Kerr-Newman-AdS black hole in
quintessential dark energy is obtained as,
\begin{equation}
T=\frac{r_{+}}{4\pi\Xi(r_{+}^{2}+a^{2})}\bigg(\frac{3r_{+}^{2}}{\ell^{2}}+\frac{a^{2}}{\ell^{2}}+1-\frac{a^{2}+Q^{2}}{r_{+}^{2}}+3\alpha\omega
r^{-1-3\omega}\bigg).
\end{equation}
The thermodynamical volume $V$ conjugated to the pressure $P$ is
given by,
\begin{equation}
V=\bigg( \frac{\partial M}{\partial P}\bigg)_{S,Q,J}.
\end{equation}
 Now we are going to study P-V criticality of corresponding black hole for three
different values of $\omega$. In the first case we consider
 $\omega=-\frac{1}{3}$.
Using the expressions (4)  and (8) and the fact that $\Delta_{r_{+}}
=0$, one can obtain a generalized Smarr formula for Kerr-Newman-AdS
black holes with quintessence as follows,
\begin{equation}
M=\bigg(\frac{\pi}{4S}\bigg\{\frac{4SJ^{2}}{\pi
\ell^{2}}+4J^{2}+\bigg[\frac{S^{2}}{\pi^{2}\ell^{2}}+\frac{S}{\pi}+Q^{2}-\frac{\alpha
S}{\pi}\bigg]^{2}\bigg\}\bigg)^{\frac{1}{2}}.
\end{equation}
The thermodynamical volume is,
\begin{equation}
V=\frac{2\pi}{3r_{+}\Xi^{2}}\bigg((r_{+}^{2}+a^{2})(2r_{+}^{2}-\frac{r_{+}^{2}a^{2}}{\ell^{2}}+a^{2}(1-\alpha)
)+Q^{2}a^{2}\bigg)+\frac{4\pi
a^{2}r_{+}^{3}}{3}\frac{\frac{2r_{+}^{4}}{\ell^{4}}+\frac{2r_{+}^{2}}{\ell^{2}}-\alpha
-\frac{\alpha r_{+}^{2}
}{\ell^{2}}}{\frac{r_{+}^{4}}{\ell^{2}}+r_{+}^{2}(1-\alpha)+Q^{2}}.
\end{equation}
The specific volume of the corresponding fluid which is related to
thermodynamical volume ($\upsilon=(\frac{3V}{4\pi})^{\frac{1}{3}}$)
will be as,
\begin{equation}
\upsilon=2r_{+}+\frac{12J^{2}(8\pi
Pr_{+}^{4}+r_{+}^{2}(3-\alpha)+Q^{2})}{r_{+}(8\pi
Pr_{+}^{4}+3r_{+}^{2}(1-\alpha)+3Q^{2})^{2}}+\frac{16J^{2}r_{+}^{3}(64\pi^{2}P^{2}r_{+}^{4}+12\pi
Pr_{+}^{2}(2-\alpha)-\frac{9\alpha}{2}}{(8\pi
Pr_{+}^{4}+3r_{+}^{2}(1-\alpha)+3Q^{2})^{3}}.
\end{equation}
The equation of state is given by,

\begin{equation}
P=\frac{T}{\upsilon}+\frac{(\alpha-1)}{2\pi
\upsilon^{2}}+\frac{2Q^{2}}{\pi
\upsilon^{4}}+\frac{48J^{2}}{\pi\upsilon^{6}}+\frac{6J^{2}B}{\pi\upsilon^{6}(\pi
T\upsilon^{3}+(1-\alpha)\upsilon^{2}+8Q^{2})^{2}},
\end{equation}
where
\begin{equation}
B=4\upsilon^{2}(\upsilon^{2}+4\alpha\upsilon^{2}-2\alpha^{2}\upsilon^{2})+16Q^{2}(9\alpha\upsilon^{2}-9\upsilon^{2}-30Q^{2})+4\pi
T\upsilon^{3}(3\upsilon^{2}(1+\alpha)-28Q^{2}-4\pi T\upsilon^{3}).
\end{equation}
We depict the $P-\upsilon$ diagram in Fig. 1 and see the
qualitative behavior is similar to the Van der Waals fluid. Critical
points occur at stationary points of inflection in the $P -
\upsilon$ diagram, where
\begin{equation}
\frac{\partial P}{\partial\upsilon}=0,~~~~~~\frac{\partial^{2}
P}{\partial\upsilon^{2}}=0,
\end{equation}
which leads to the following equation,
\begin{equation}
\upsilon_{c}=2\sqrt{\frac{3(Q^{2}+\sqrt{Q^{4}-10\alpha
J^{2}+10J^{2}})}{(1-\alpha)}}.
\end{equation}
Due to the complexity of the obtained relations for temperature and
pressure, one can determine critical points numerically. We obtain
the critical specific volume $\upsilon_{c}$, critical temperature
$T_{c}$, critical pressure $P_{c}$ and the universal critical ratio
($\frac{P_{c}\upsilon_{c}}{T_{c}}$) for different values of $Q$, $J$
and $\alpha$ as follows,

\begin{center}
\begin{tabular}{|c|c|c|c|c|c|c|}
  \hline
$\alpha$\quad & $J$\quad & $Q$\quad & $\upsilon_{c}$\quad & $T_{c}$\quad & $P_{c}$\quad & $\frac{P_{c}\upsilon_{c}}{T_{c}}$\quad \\
\hline
0.1\quad & 0.1\quad & 0.05\quad & 2.0083\quad & 0.1138\quad & 0.0235\quad & 0.4160\quad \\
\hline
0.5\quad & 0.1\quad & 0.05\quad & 2.3295\quad & 0.0544\quad & 0.0544\quad & 0.4158\quad \\
\hline
0.1\quad & 0.2\quad & 0.05\quad & 2.8343\quad & 0.0807\quad & 0.0097\quad & 0.4163\quad \\
\hline
0.1\quad & 0.1\quad & 0.2\quad & 2.1374\quad & 0.0103\quad & 0.0118\quad & 0.4082\quad \\
\hline
\end{tabular}
\\[0pt]
Table $1$: $\upsilon_{c}$, $T_{c}$, $P_{c}$ and their ratio for
different value of $Q$, $J$ and $\alpha$.
\end{center}
As we see the critical specific volume and critical pressure are
increased and critical temperature is decreased by increasing
$\alpha$. When angular momentum increases $\upsilon_{c}$ and $T_{c}$
increase and $P_{c}$ decreases. But by increasing charge of black
hole only $\upsilon_{c}$ increases and both $T_{c}$ and $P_{c}$
decrease. Also table (1) shows that
$\frac{P_{c}\upsilon_{c}}{T_{c}}$ is increased
by increasing $J$ and is decreased by increasing $\alpha$ and $Q$.\\
\begin{figure}
\hspace*{1cm}
\begin{center}
\epsfig{file=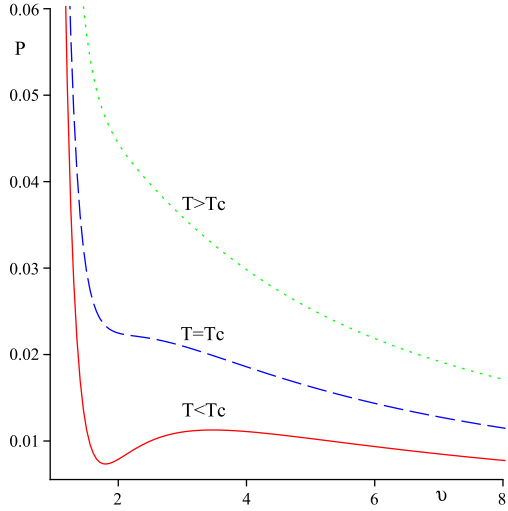,width=7cm} \caption{\small{The $P -\upsilon $
diagram for $Q = 0.05$, $J = 0.1$, $\alpha=0.1$ and different values
of $T$.}}
\end{center}
\end{figure}
 The second case we consider $\omega=-\frac{2}{3}$.
The generalized Smarr formula for this case is obtained as,
\begin{equation}
M=\bigg(\frac{\pi}{4S}\bigg\{\frac{4SJ^{2}}{\pi
\ell^{2}}+4J^{2}+\bigg[\frac{S^{2}}{\pi^{2}\ell^{2}}+\frac{S}{\pi}+Q^{2}-\alpha
\bigg(\frac{S}{\pi}\bigg)^{\frac{3}{2}}\bigg]^{2}\bigg\}\bigg)^{\frac{1}{2}}.
\end{equation}
The thermodynamical volume is determined as,
\begin{equation}
V=\frac{2\pi}{3r_{+}\Xi^{2}}\bigg((r_{+}^{2}+a^{2})(2r_{+}^{2}-\frac{r_{+}^{2}a^{2}}{\ell^{2}}+a^{2}-\alpha
r_{+}a^{2})+Q^{2}a^{2}\bigg)-\frac{4\pi
a^{2}r_{+}^{3}}{3}\frac{-\frac{2r_{+}^{4}}{\ell^{4}}-\frac{2r_{+}^{2}}{\ell^{2}}+\frac{3}{2}\alpha
r_{+}+\frac{3\alpha r_{+}^{3}
}{2\ell^{2}}}{\frac{r_{+}^{4}}{\ell^{2}}+r_{+}^{2}+Q^{2}-\alpha
r_{+}^{3}}.
\end{equation}
The specific volume is obtained as,
\begin{equation}
\upsilon=2r_{+}+\frac{12J^{2}(8\pi Pr_{+}^{4}+3r_{+}^{2}-\alpha
r_{+}^{3}+Q^{2})}{r_{+}(8\pi Pr_{+}^{4}+3r_{+}^{2}-3\alpha
r_{+}^{3}+3Q^{2})^{2}}+\frac{8J^{2}r_{+}^{3}(128\pi^{2}P^{2}r_{+}^{4}+48\pi
Pr_{+}^{2}-\frac{27}{2}\alpha r_{+}-36\alpha\pi P r_{+}^{3})}{(8\pi
Pr_{+}^{4}+3r_{+}^{2}-3\alpha r_{+}^{3}+3Q^{2})^{3}}.
\end{equation}
By using equations (5), (10) and (21), we can obtain  the equation
of state as,

\begin{equation}
P=\frac{T}{\upsilon}-\frac{1}{2\pi \upsilon^{2}}+\frac{2Q^{2}}{\pi
\upsilon^{4}}+\frac{\alpha}{2\pi\upsilon}+\frac{48J^{2}}{\pi\upsilon^{6}}+\frac{6J^{2}B}{\pi\upsilon^{6}(\pi
T\upsilon^{3}+\upsilon^{2}+8Q^{2}-\frac{\alpha\upsilon^{3}}{4})^{2}},
\end{equation}
where
\begin{equation}
B=\upsilon^{2}(4\upsilon^{2}-2\alpha\upsilon^{3}+\frac{\alpha^{2}\upsilon^{4}}{2})+4Q^{2}(13\alpha\upsilon^{3}-36\upsilon^{2}-112Q^{2})+4\pi
T\upsilon^{3}(3\upsilon^{2}-28Q^{2}-4\pi T\upsilon^{3}).
\end{equation}
We have plotted $P-\upsilon$ diagram in Fig. 2 which shows the Van
der Waals like behavior. By using equations (17) and (22) the
critical specific volume is obtained by,
\begin{equation}
\upsilon_{c}=2\sqrt{3(Q^{2}+\sqrt{Q^{4}+20J^{2}})}.
\end{equation}
We obtain critical quantities and ratio of them for different values
of $Q$, $J$ and $\alpha$ as follows,
\begin{center}
\begin{tabular}{|c|c|c|c|c|c|c|}
  \hline
$\alpha$\quad & $J$\quad & $Q$\quad & $\upsilon_{c}$\quad & $T_{c}$\quad & $P_{c}$\quad & $\frac{P_{c}\upsilon_{c}}{T_{c}}$\quad \\
\hline
0.1\quad & 0.1\quad & 0.05\quad & 2.3230\quad & 0.1071\quad & 0.0244\quad & 0.5309\quad \\
\hline
0.3\quad & 0.1\quad & 0.05\quad & 1.9557\quad & 0.1299\quad & 0.0520\quad & 0.7837\quad \\
\hline
0.1\quad & 0.2\quad & 0.05\quad & 3.2807\quad & 0.07131\quad & 0.0123\quad & 0.5658\quad \\
\hline
0.1\quad & 0.1\quad & 0.2\quad & 2.4223\quad & 0.0973\quad & 0.0211\quad & 0.5257\quad \\
\hline
\end{tabular}
\\[0pt]
Table $2$: $\upsilon_{c}$, $T_{c}$, $P_{c}$ and their ratio for
different value of $Q$, $J$ and $\alpha$.
\end{center}
Table (2) shows that $T_{c}$ and $P_{c}$ are increased by increasing
$\alpha$ and decreasing $J$ but they are decreased by increasing
$Q$. As we see $\upsilon_{c}$ increases when $J$ and $Q$ increase
and it decreases when $\alpha$ increases. Also we find that
$\frac{P_{c}\upsilon_{c}}{T_{c}}$ is increased by increasing
$\alpha$ and $J$ and is decreased by increasing $Q$.
\begin{figure}
\hspace*{1cm}
\begin{center}
\epsfig{file=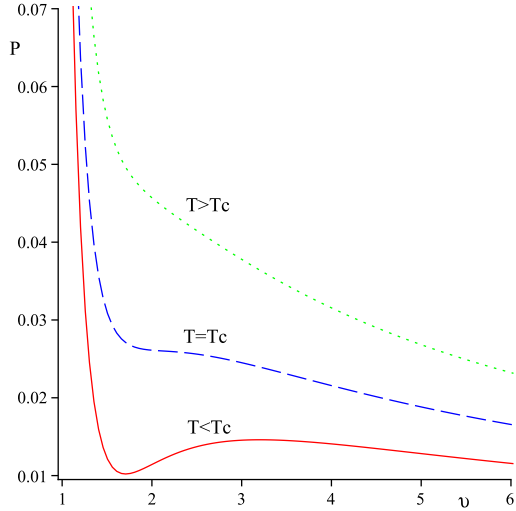,width=7cm} \caption{\small{The $P -\upsilon $
diagram for $Q = 0.05$, $J = 0.1$, $\alpha=0.1$ and different values
of $T$.}}
\end{center}
\end{figure}

In third case, we assume $\omega=-1$. In this case the generalized
Smarr formula is,
\begin{equation}
M=\bigg(\frac{\pi}{4S}\bigg\{\frac{4SJ^{2}}{\pi
\ell^{2}}+4J^{2}+\bigg[\frac{S^{2}}{\pi^{2}\ell^{2}}+\frac{S}{\pi}+Q^{2}-\frac{\alpha
S^{2}}{\pi^{2}}\bigg]^{2}\bigg\}\bigg)^{\frac{1}{2}}.
\end{equation}
The thermodynamical volume will be as,
\begin{equation}
V=\frac{2\pi}{3r_{+}\Xi^{2}}\bigg((r_{+}^{2}+a^{2})(2r_{+}^{2}-\frac{r_{+}^{2}a^{2}}{\ell^{2}}+a^{2}-\alpha
r_{+}^{2}a^{2})+Q^{2}a^{2}\bigg)-\frac{4\pi
a^{2}r_{+}^{3}}{3}\frac{-\frac{2r_{+}^{4}}{\ell^{4}}-\frac{2r_{+}^{2}}{\ell^{2}}+2\alpha
r_{+}^{2}+\frac{2\alpha r_{+}^{4}
}{\ell^{2}}}{\frac{r_{+}^{4}}{\ell^{2}}+r_{+}^{2}+Q^{2}-\alpha
r_{+}^{4}}.
\end{equation}

The specific volume of the corresponding is fluid as fellows,
\begin{equation}
\upsilon=2r_{+}+\frac{12J^{2}(8\pi Pr_{+}^{4}+3r_{+}^{2}-\alpha
r_{+}^{4}+Q^{2})}{r_{+}(8\pi Pr_{+}^{4}+3r_{+}^{2}-3\alpha
r_{+}^{4}+3Q^{2})^{2}}+\frac{16J^{2}r_{+}^{3}(64\pi^{2}P^{2}r_{+}^{4}+24\pi
Pr_{+}^{2}-9\alpha r_{+}^{2}-24\alpha\pi P r_{+}^{4})}{(8\pi
Pr_{+}^{4}+3r_{+}^{2}-3\alpha r_{+}^{4}+3Q^{2})^{3}}.
\end{equation}

The equation of state is given by,

\begin{equation}
P=\frac{T}{\upsilon}-\frac{1}{2\pi \upsilon^{2}}+\frac{2Q^{2}}{\pi
\upsilon^{4}}+\frac{3\alpha}{8\pi}+\frac{48J^{2}}{\pi\upsilon^{6}}+\frac{6J^{2}B}{\pi\upsilon^{6}(\pi
T\upsilon^{3}+\upsilon^{2}+8Q^{2})^{2}},
\end{equation}

where
\begin{equation}
B=\upsilon^{2}(4\upsilon^{2}-\alpha\upsilon^{4}+8\alpha^{2}\upsilon^{4})+8Q^{2}(2\alpha\upsilon^{4}-18\upsilon^{2}-56Q^{2})+\pi
T\upsilon^{3}(12\upsilon^{2}-\alpha\upsilon^{4}-112Q^{2}-16\pi
T\upsilon^{3}).
\end{equation}
Figure (3) shows qualitative behavior $P$ with respect to $\upsilon$
which is same as the Van der Waals fluid. In that case we obtain
$\upsilon_{c}$ as follows,

\begin{equation}
\upsilon_{c}=2\sqrt{3(Q^{2}+\sqrt{Q^{4}+10J^{2}})}
\end{equation}
We indicate changes of $\upsilon_{c}$, $T_{c}$ and $P_{c}$ for
different values of $Q$, $J$ and $\alpha$ in table (3). As we see
$\upsilon_{c}$ and $\frac{P_{c}\upsilon_{c}}{T_{c}}$ are increased
but $T_{c}$ and $P_{c}$ are decreased by increasing $Q$ and $J$.
Also we observe when $\alpha$ increases $\upsilon_{c}$ and $T_{c}$
don't change  but $P_{c}$ and $\frac{P_{c}\upsilon_{c}}{T_{c}}$
increase. By comparing results we find that $P_{c}$ and
$\frac{P_{c}\upsilon_{c}}{T_{c}}$ decreases when $\omega$ increases
from $-1$ to $-\frac{1}{3}$. We find that for $\omega=-1$ and
$-\frac{2}{3}$ and for $\alpha>0.4$,
$\frac{P_{c}\upsilon_{c}}{T_{c}}$ will be greater than one. Also we
notice that the obtained result for $\omega=-\frac{1}{3}$ is very
similar to case of without dark energy parameter as Ref. [14].

\begin{center}
\begin{tabular}{|c|c|c|c|c|c|c|}
  \hline
$\alpha$\quad & $J$\quad & $Q$\quad & $\upsilon_{c}$\quad & $T_{c}$\quad & $P_{c}$\quad & $\frac{P_{c}\upsilon_{c}}{T_{c}}$\quad \\
\hline
0.1\quad & 0.1\quad & 0.05\quad & 1.9557\quad & 0.1299\quad & 0.0357\quad & 0.5386\quad \\
\hline
0.3\quad & 0.1\quad & 0.05\quad & 1.9557\quad & 0.1299\quad & 0.0520\quad & 0.7837\quad \\
\hline
0.1\quad & 0.2\quad & 0.05\quad & 2.7603\quad & 0.0921\quad & 0.0196\quad & 0.5891\quad \\
\hline
0.1\quad & 0.1\quad & 0.2\quad & 2.0748\quad & 0.1182\quad & 0.0309\quad & 0.5433\quad \\
\hline
\end{tabular}
\\[0pt]
Table $3$: $\upsilon_{c}$, $T_{c}$, $P_{c}$ and their ratio for
different values of $Q$, $J$ and $\alpha$.
\end{center}

\begin{figure}
\hspace*{1cm}
\begin{center}
\epsfig{file=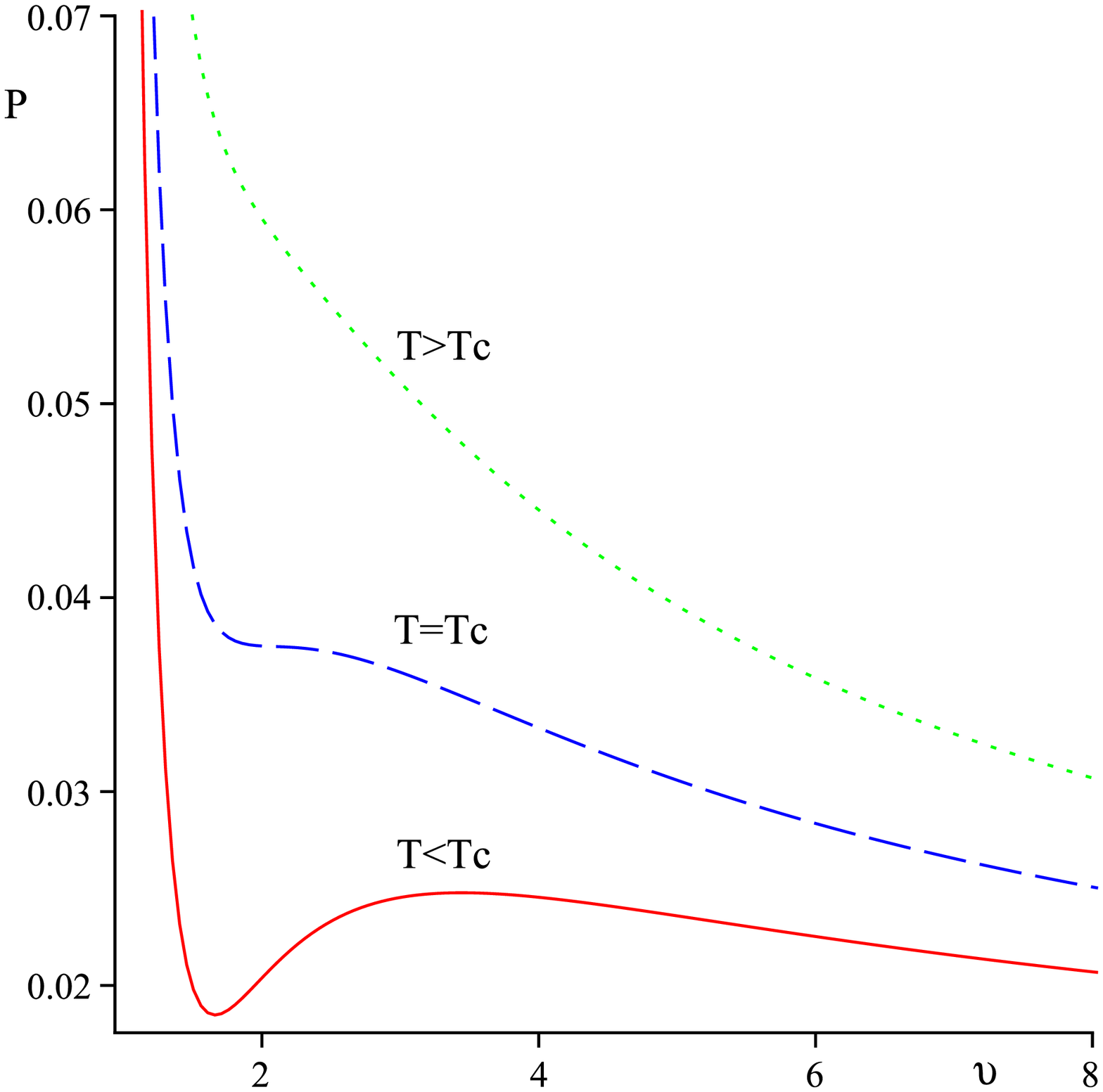,width=7cm} \caption{\small{The $P -\upsilon $
diagram for $Q = 0.05$, $J = 0.1$, $\alpha=0.1$ and different values
of $T$.}}
\end{center}
\end{figure}

\begin{figure}
\hspace*{1cm}
\begin{center}
\epsfig{file=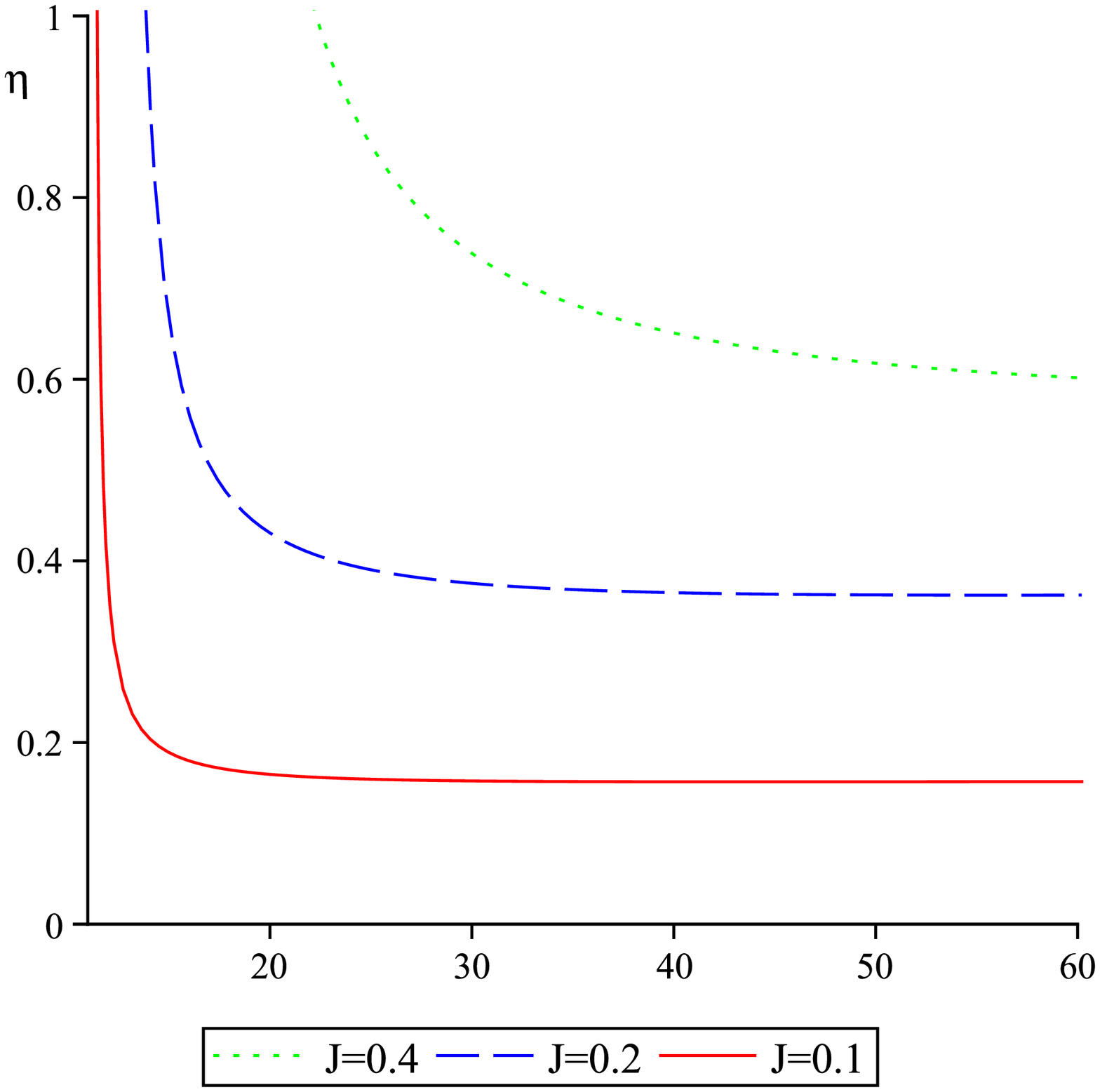,width=7cm}
\epsfig{file=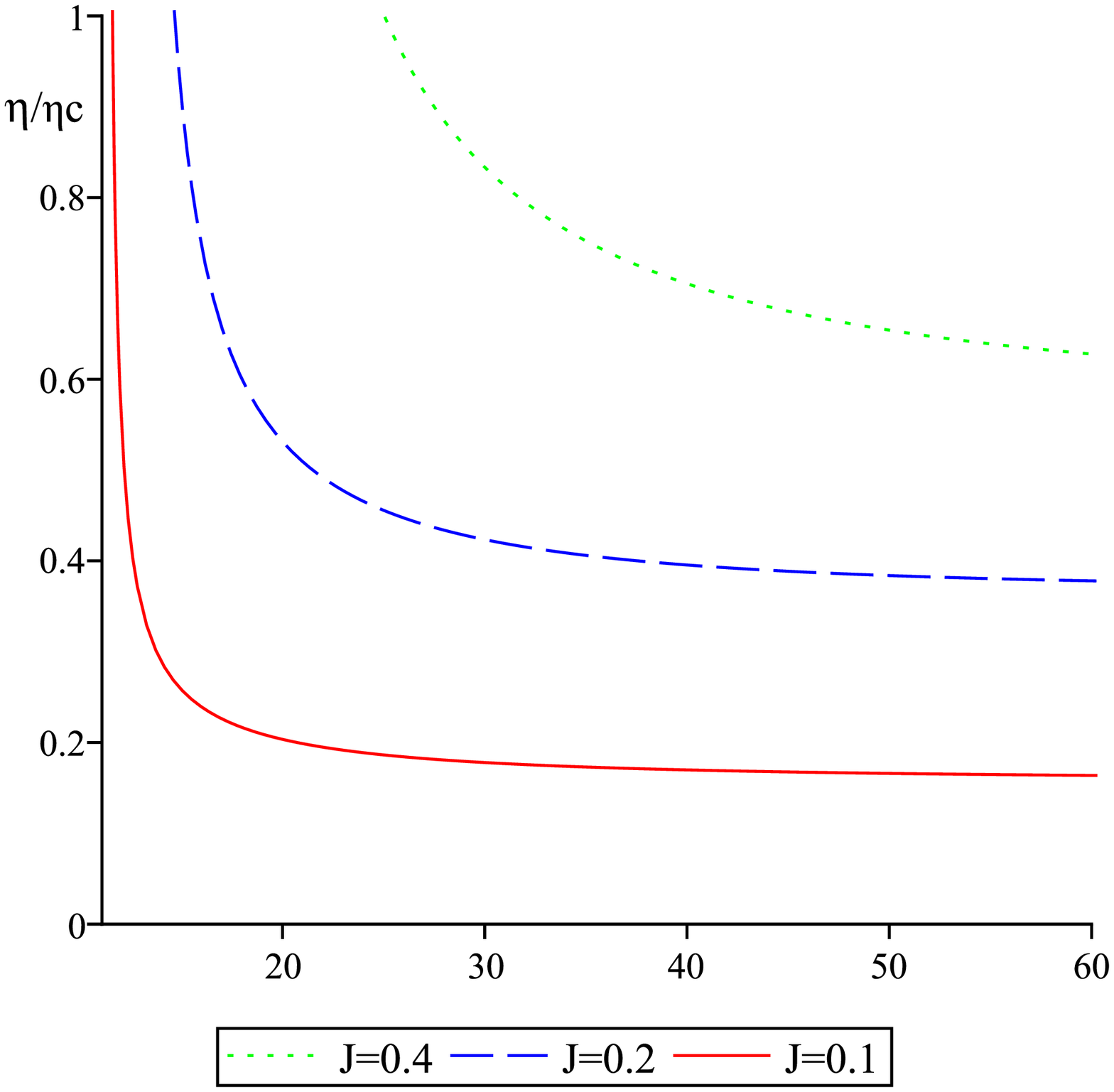,width=7cm}\caption{\small{Left plot: $\eta$
 with respect to $V_{R}$ for $\alpha=0.1$, $Q=0.05$, $P_{T}=0.07$, $P_{B}=0.06$, $V_{L}=10$ and different values of $J$ ; Right
 plot: $\frac{\eta}{\eta_{C}}$ with respect to $V_{R}$ for $\alpha=0.1$, $Q=0.05$, $P_{T}=0.07$, $P_{B}=0.06$, $V_{L}=10$ and different values of $J$
}}
\end{center}
\end{figure}
\begin{figure}
\hspace*{1cm}
\begin{center}
\epsfig{file=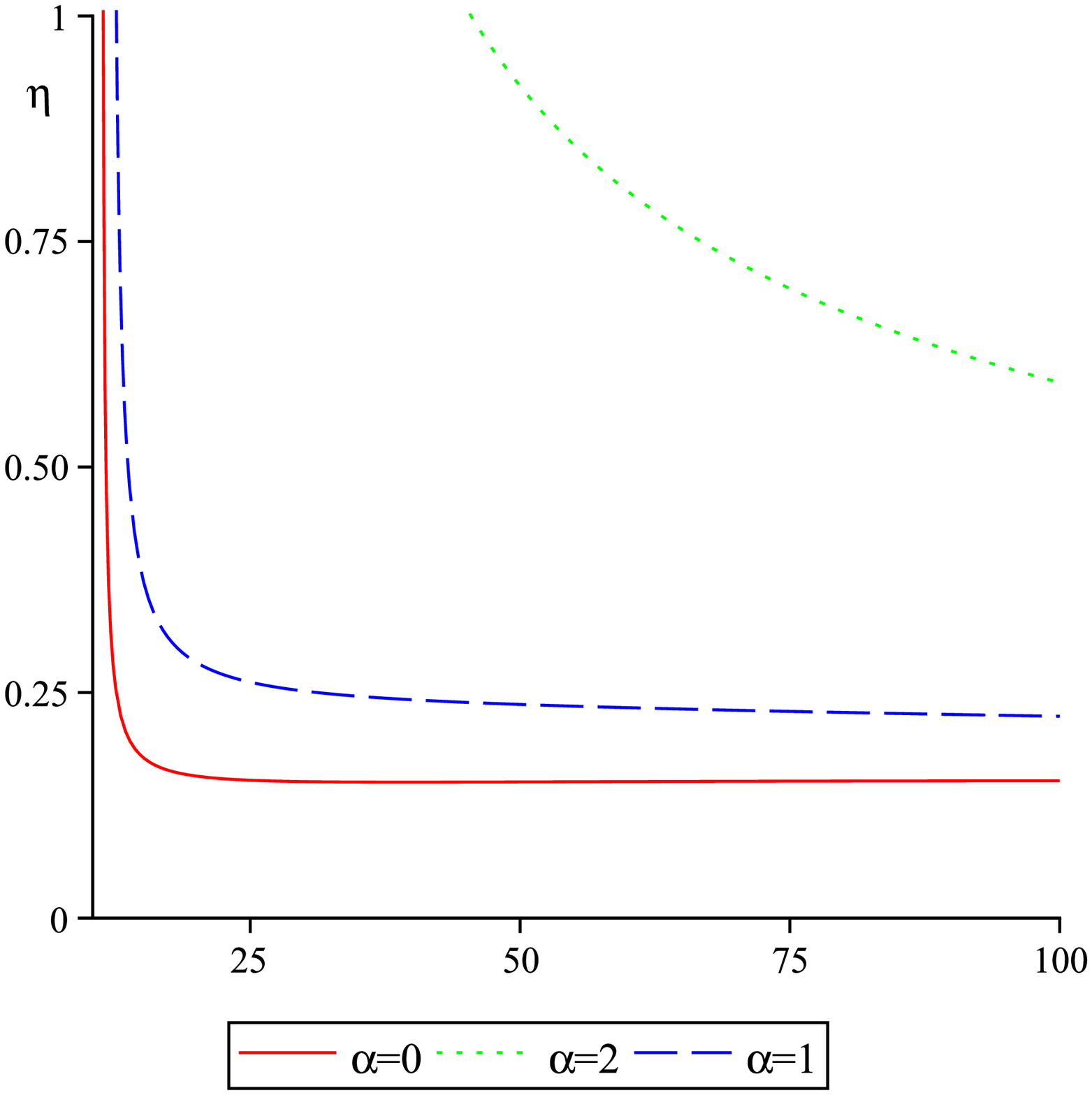,width=7cm}
\epsfig{file=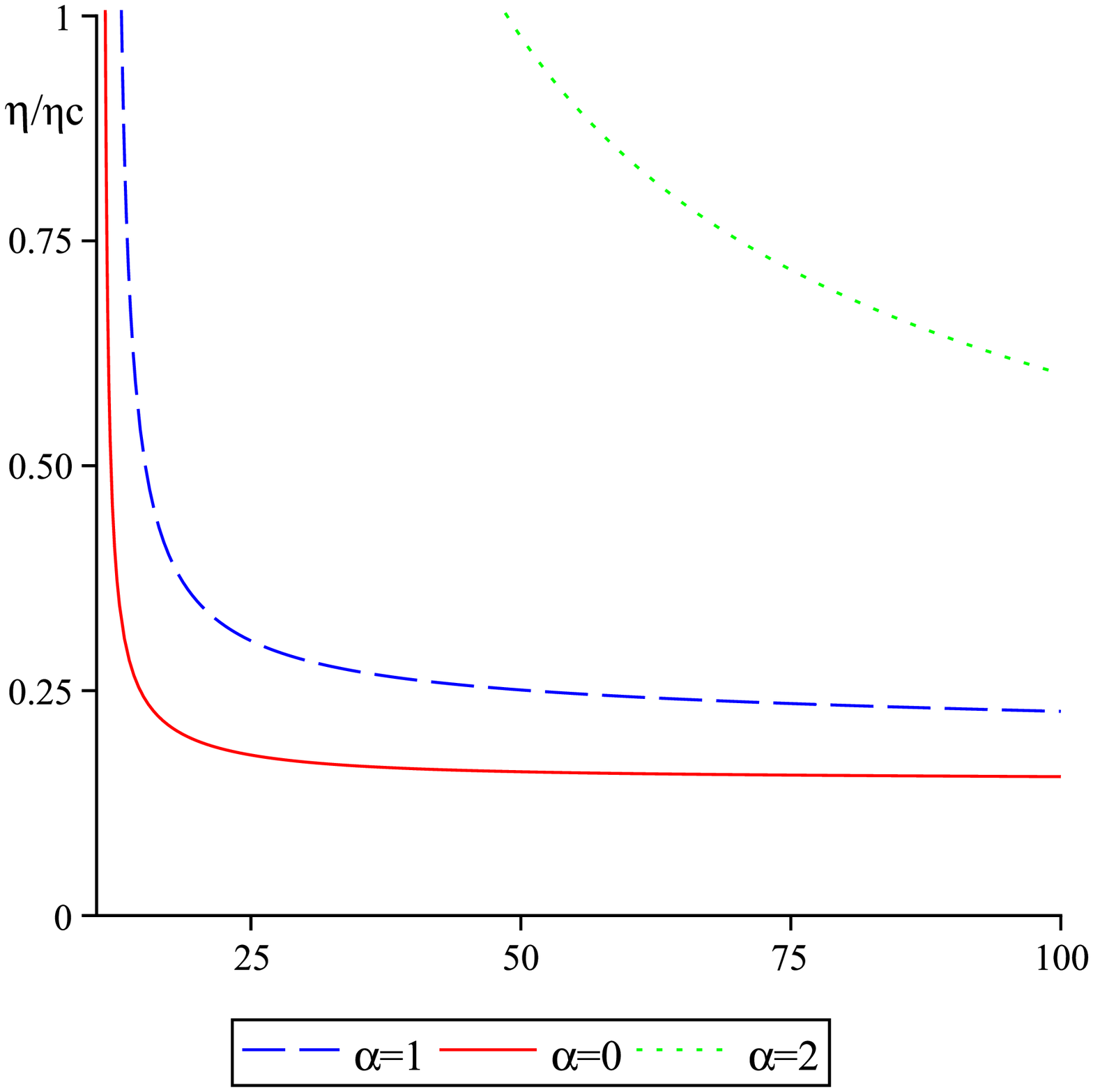,width=7cm}\caption{\small{Left plot: $\eta$
 with respect to $V_{R}$ for $J=0.1$, $Q=0.05$, $P_{T}=0.07$, $P_{B}=0.06$, $V_{L}=10$ and different values of $\alpha$ ; Right
 plot: $\frac{\eta}{\eta_{C}}$ with respect to $V_{R}$ for $J=0.1$, $Q=0.05$, $P_{T}=0.07$, $P_{B}=0.06$, $V_{L}=10$ and different values of $\alpha$
}}
\end{center}
\end{figure}
\begin{figure}
\hspace*{1cm}
\begin{center}
\epsfig{file=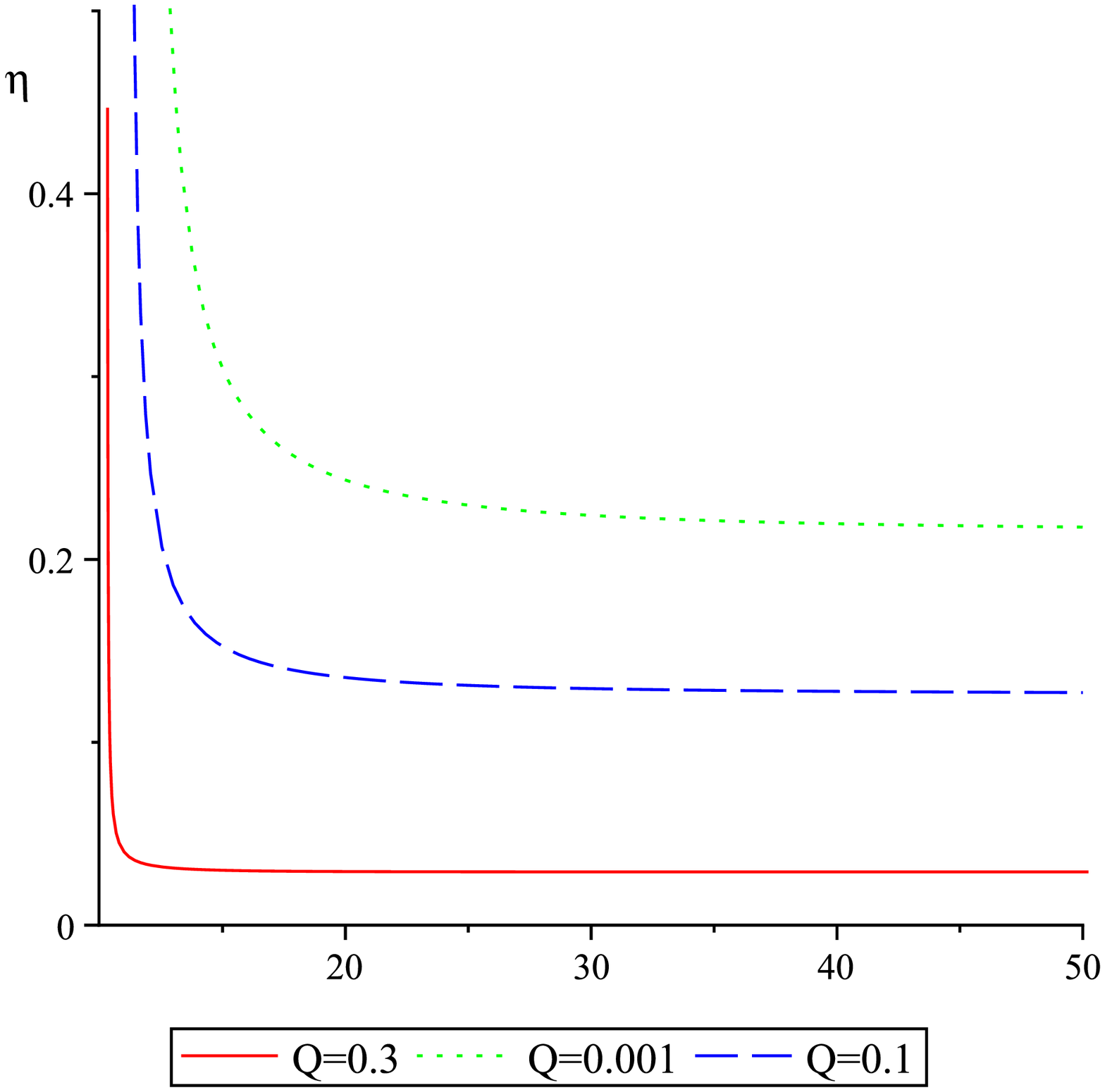,width=7cm}
\epsfig{file=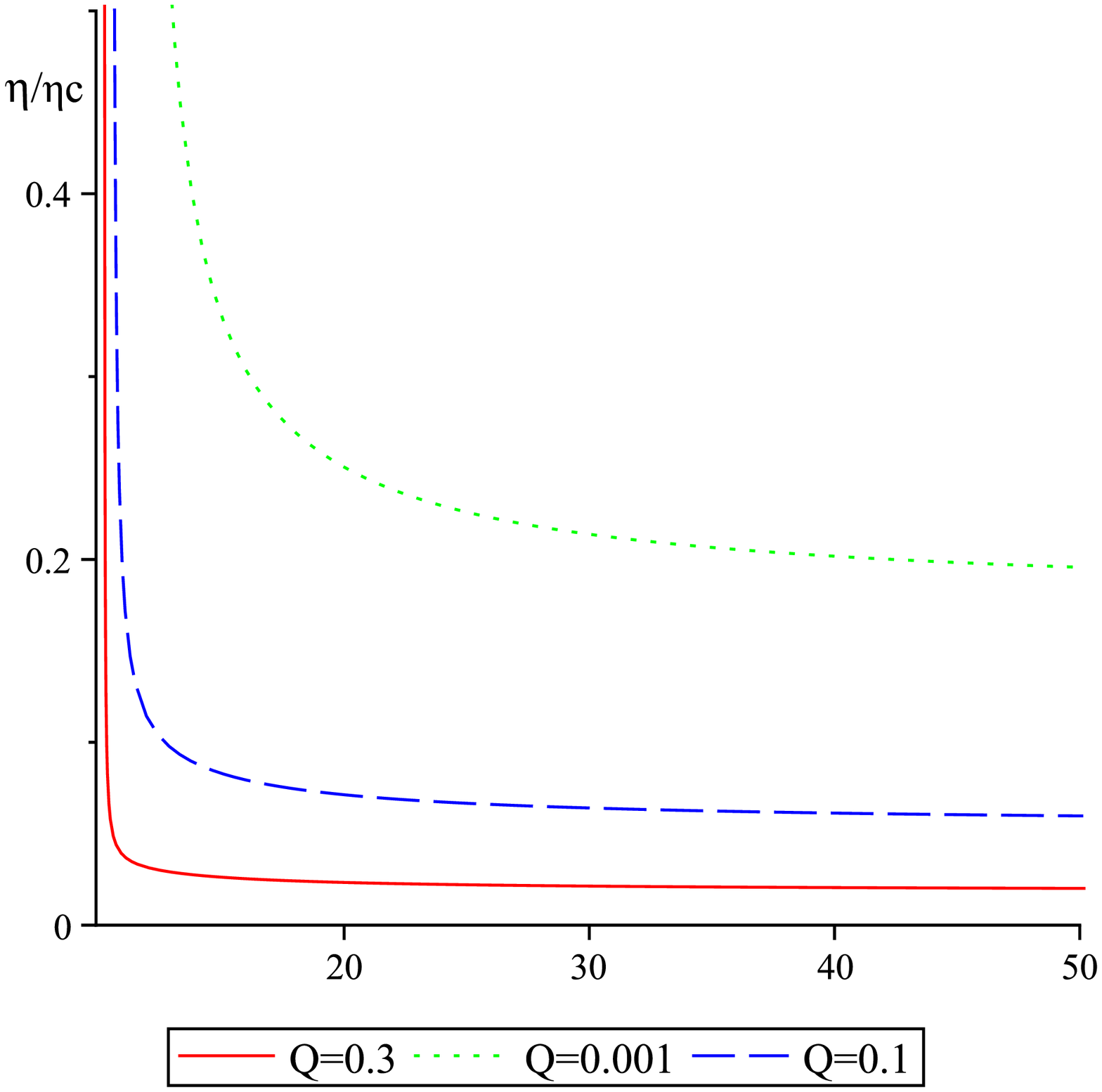,width=7cm}\caption{\small{Left plot: $\eta$
 with respect to $V_{R}$ for $\alpha=0.1$, $J=0.1$, $P_{T}=0.07$, $P_{B}=0.06$, $V_{L}=10$ and different values of $Q$ ; Right
 plot: $\frac{\eta}{\eta_{C}}$ with respect to $V_{R}$ for $\alpha=0.1$, $J=0.1$, $P_{T}=0.07$, $P_{B}=0.06$, $V_{L}=10$ and different values of $Q$
}}
\end{center}
\end{figure}
\begin{figure}
\hspace*{1cm}
\begin{center}
\epsfig{file=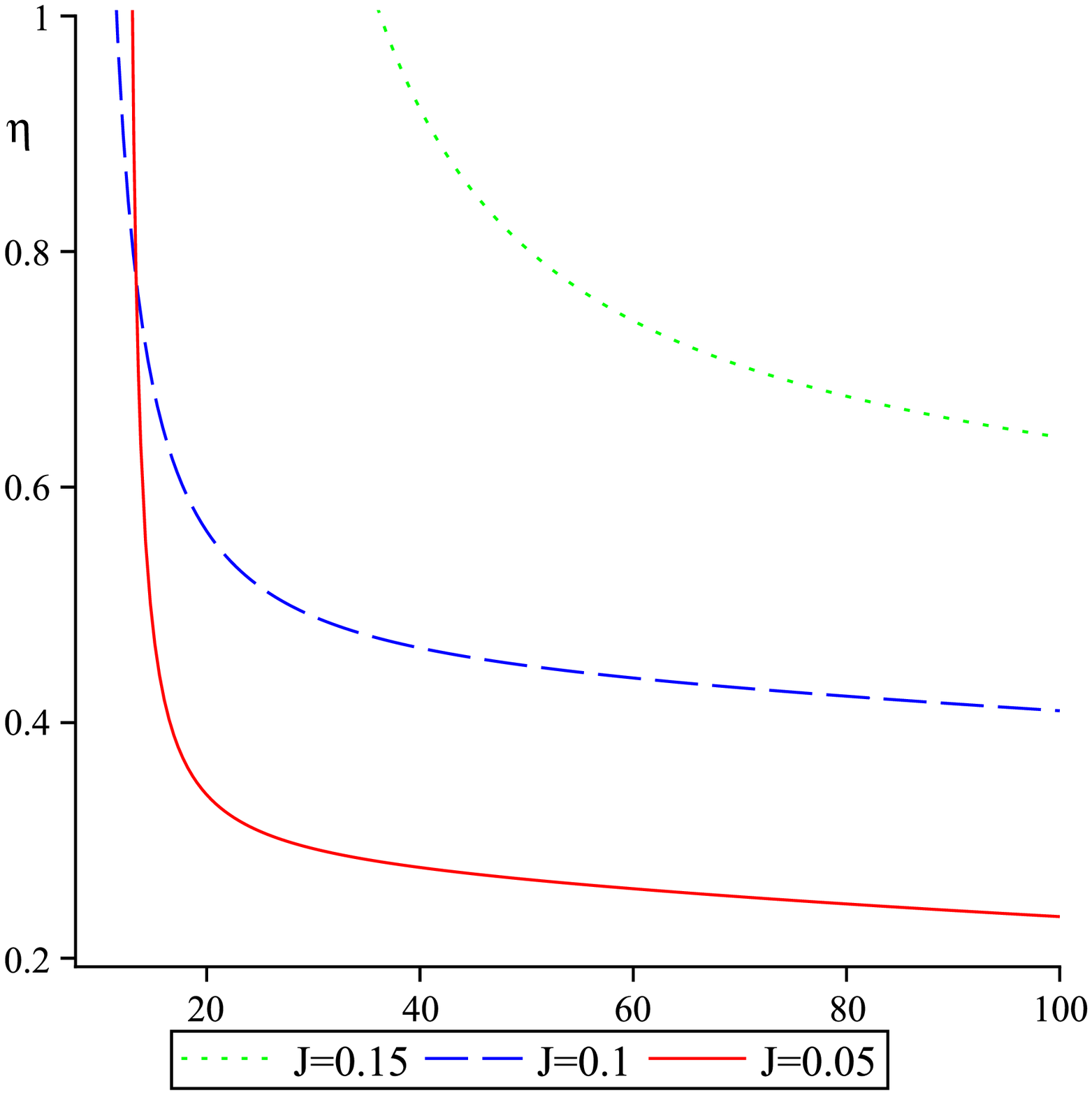,width=7cm}
\epsfig{file=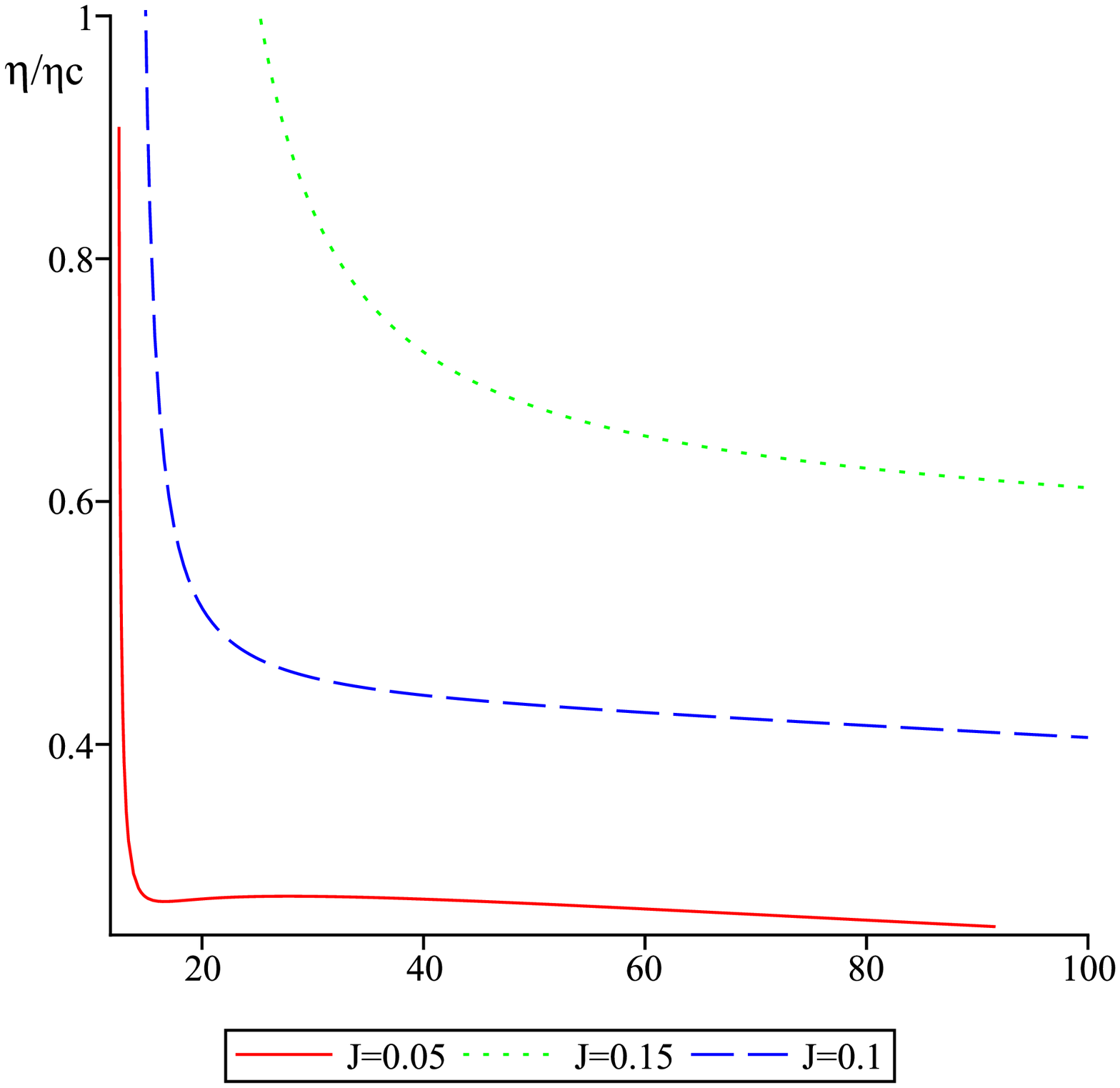,width=7cm}\caption{\small{Left plot: $\eta$
 with respect to $V_{R}$ for $\alpha=0.1$, $Q=0.05$, $P_{T}=0.07$, $P_{B}=0.06$, $V_{L}=10$ and different values of $J$ ; Right
 plot: $\frac{\eta}{\eta_{C}}$ with respect to $V_{R}$ for $\alpha=0.1$, $Q=0.05$, $P_{T}=0.07$, $P_{B}=0.06$, $V_{L}=10$ and different values of $J$
}}
\end{center}
\end{figure}
\begin{figure}
\hspace*{1cm}
\begin{center}
\epsfig{file=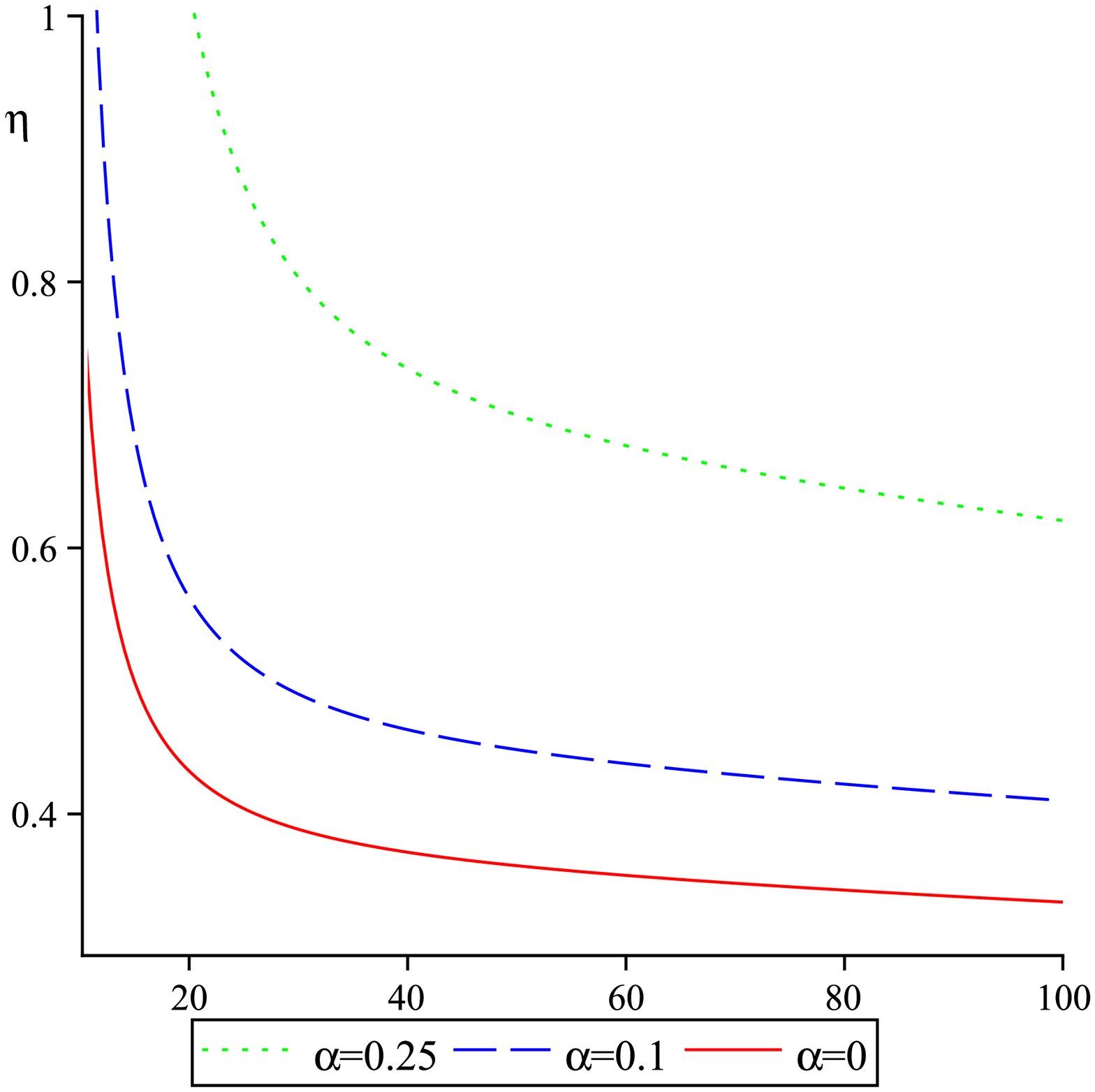,width=7cm}
\epsfig{file=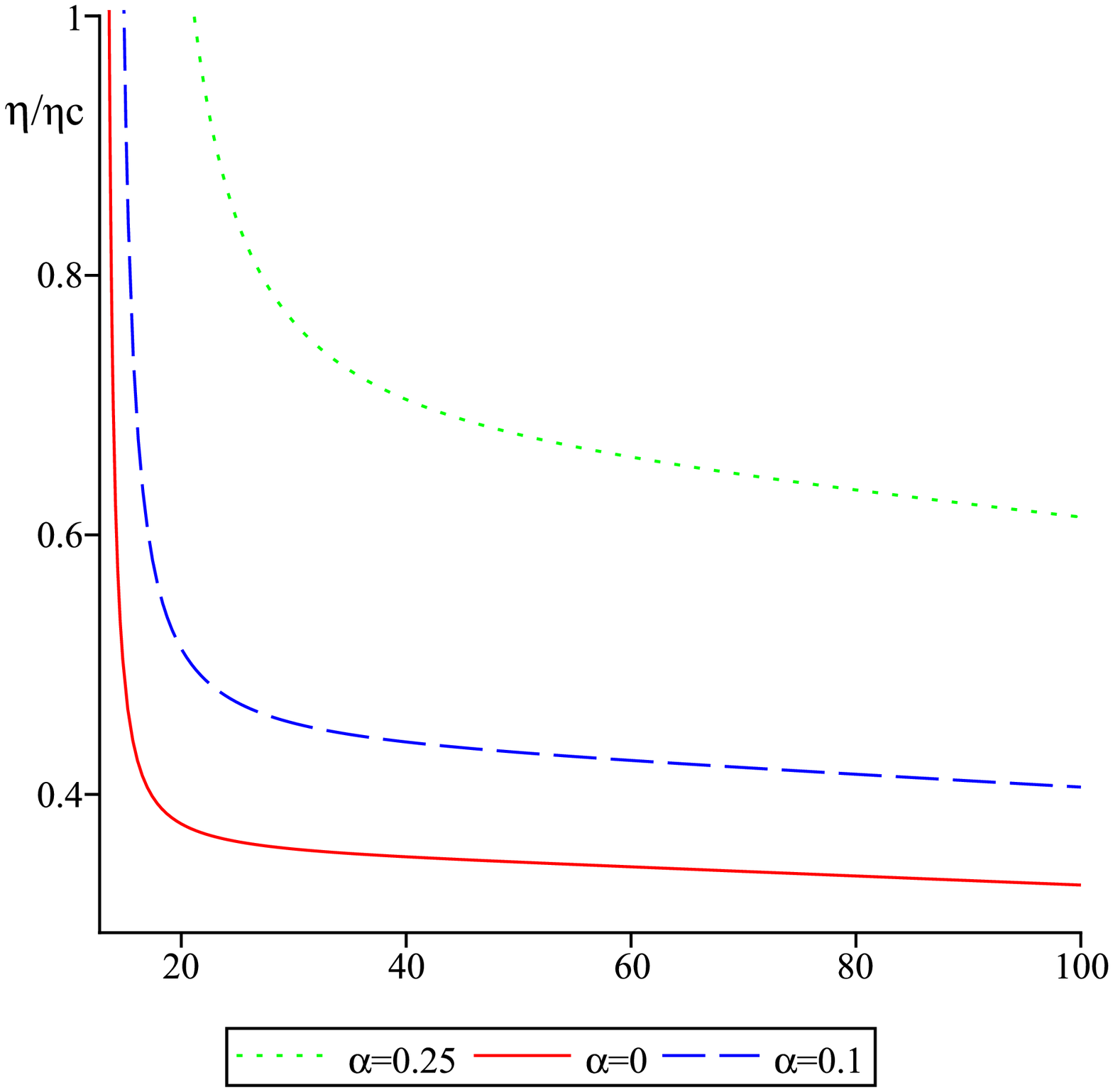,width=7cm}\caption{\small{Left plot: $\eta$
 with respect to $V_{R}$ for $J=0.1$, $Q=0.05$, $P_{T}=0.07$, $P_{B}=0.06$, $V_{L}=10$ and different values of $\alpha$ ; Right
 plot: $\frac{\eta}{\eta_{C}}$ with respect to $V_{R}$ for $J=0.1$, $Q=0.05$, $P_{T}=0.07$, $P_{B}=0.06$, $V_{L}=10$ and different values of $\alpha$
}}
\end{center}
\end{figure}
\begin{figure}
\hspace*{1cm}
\begin{center}
\epsfig{file=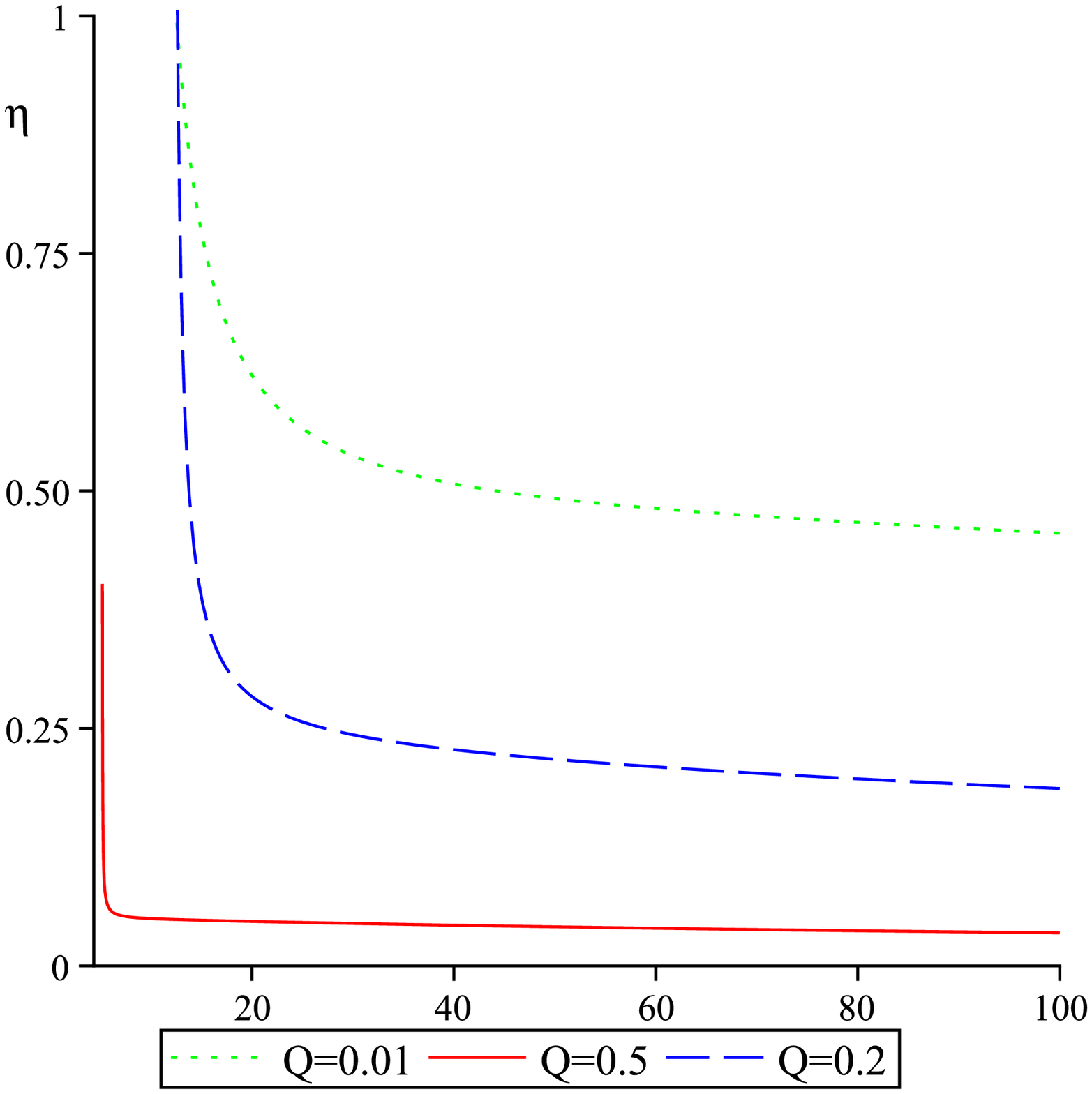,width=7cm}
\epsfig{file=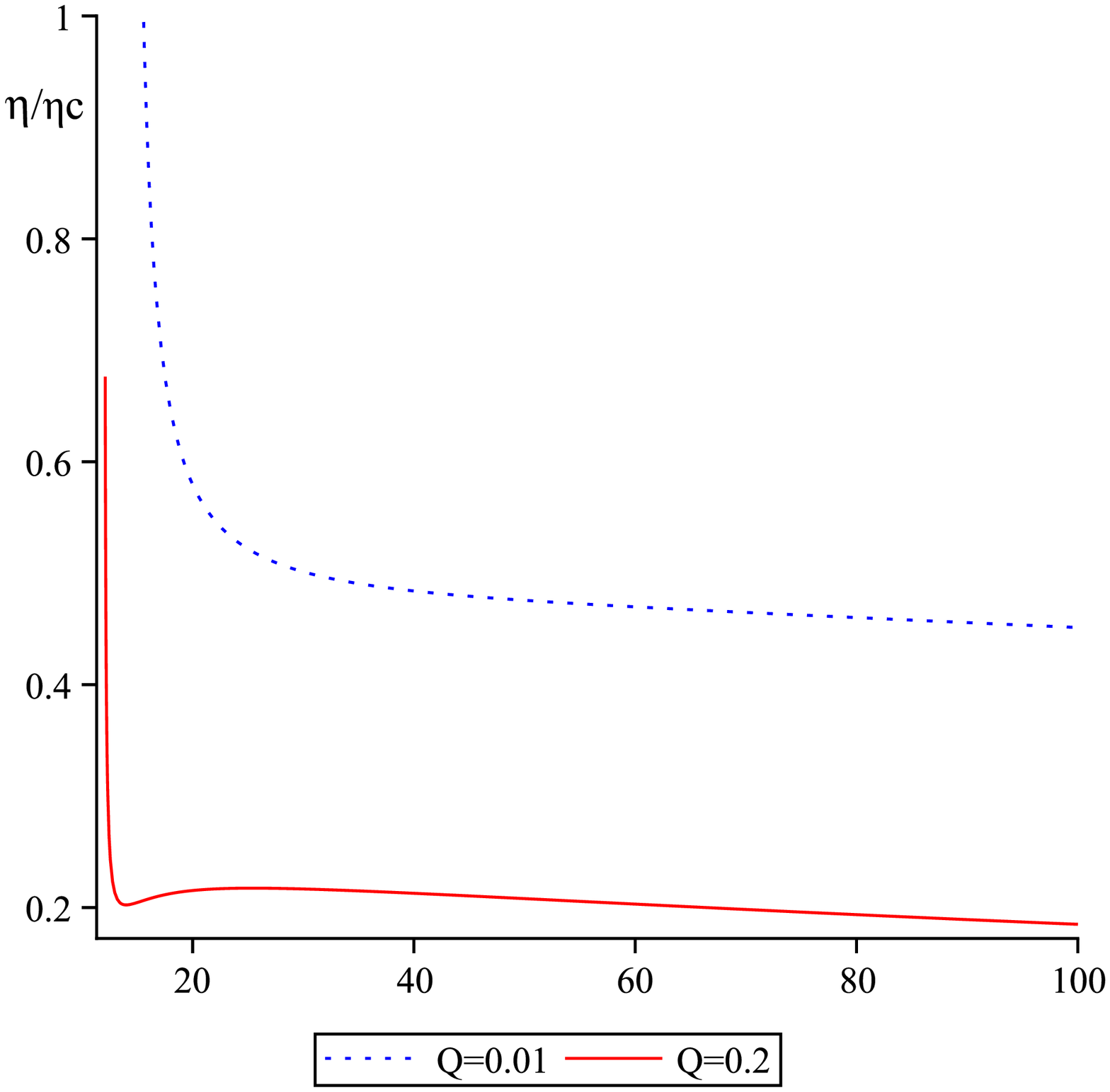,width=7cm}\caption{\small{Left plot: $\eta$
 with respect to $V_{R}$ for $\alpha=0.1$, $J=0.1$, $P_{T}=0.07$, $P_{B}=0.06$, $V_{L}=10$ and different values of $Q$ ; Right
 plot: $\frac{\eta}{\eta_{C}}$ with respect to $V_{R}$ for $\alpha=0.1$, $J=0.1$, $P_{T}=0.07$, $P_{B}=0.06$, $V_{L}=10$ and different values of $Q$
}}
\end{center}
\end{figure}

\begin{figure}
\hspace*{1cm}
\begin{center}
\epsfig{file=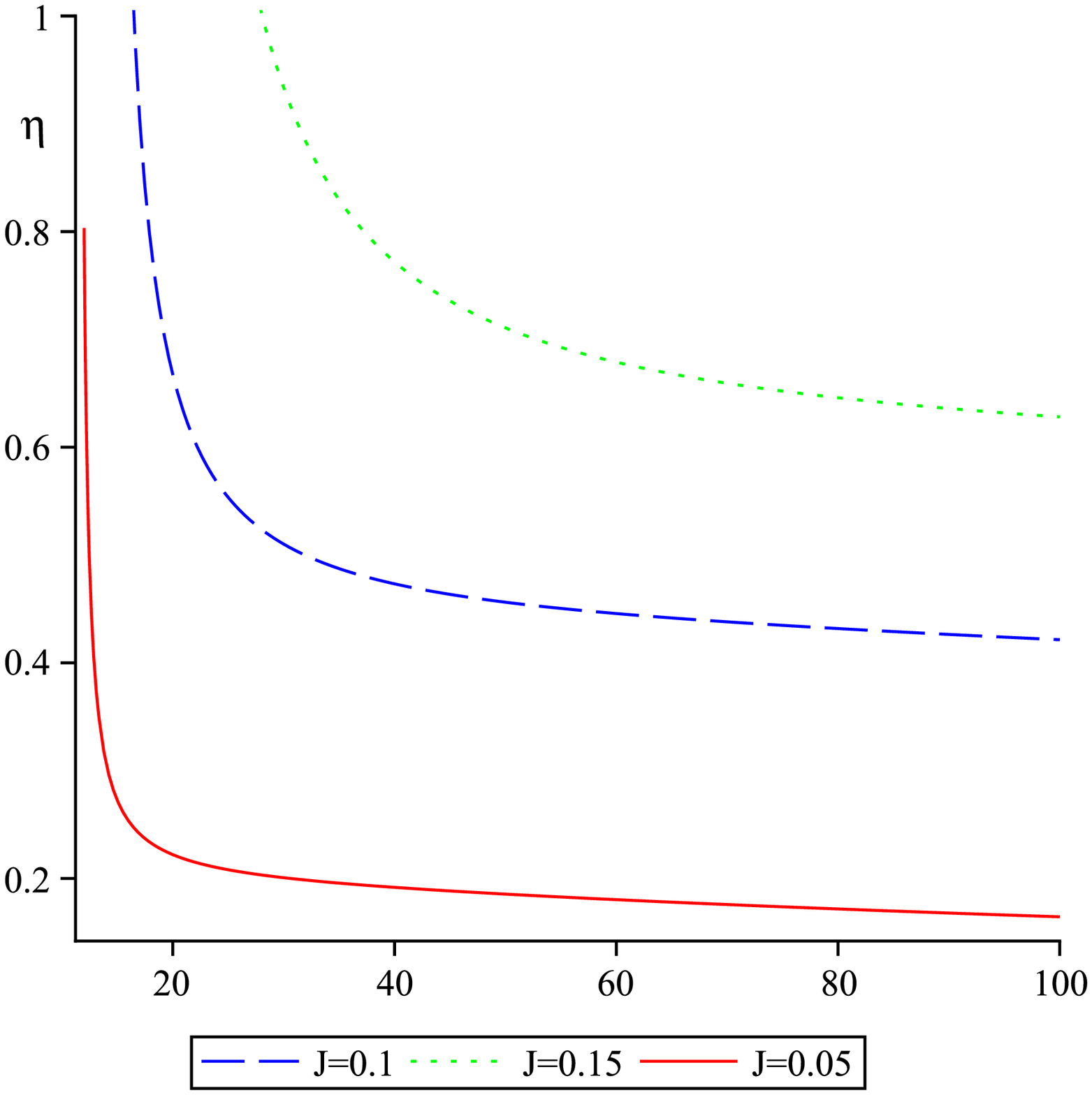,width=7cm}
\epsfig{file=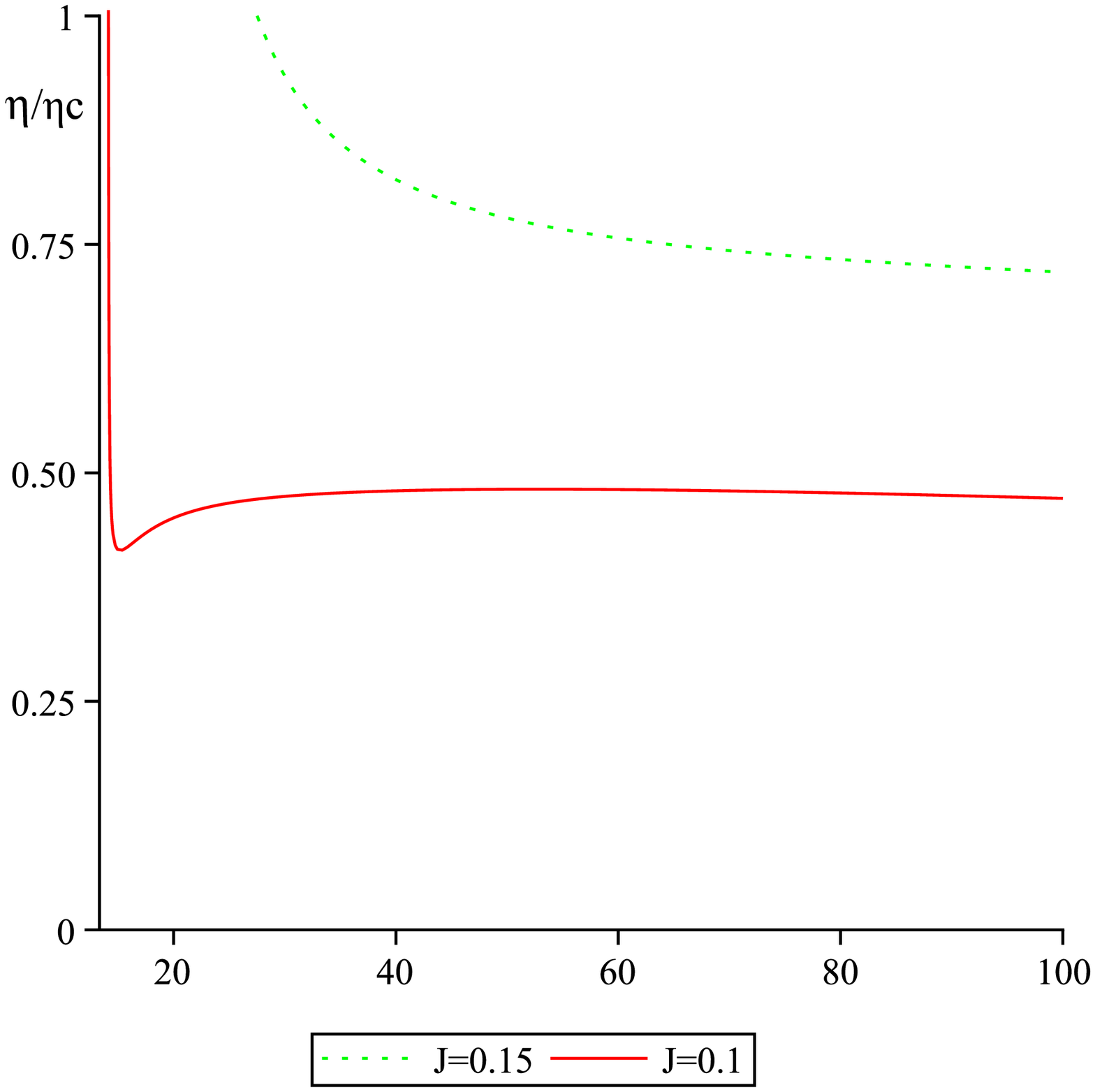,width=7cm}\caption{\small{Left plot: $\eta$
 with respect to $V_{R}$ for $\alpha=0.1$, $Q=0.05$, $P_{T}=0.07$, $P_{B}=0.06$, $V_{L}=10$ and different values of $J$ ; Right
 plot: $\frac{\eta}{\eta_{C}}$ with respect to $V_{R}$ for $\alpha=0.1$, $Q=0.05$, $P_{T}=0.07$, $P_{B}=0.06$, $V_{L}=10$ and different values of $J$
}}
\end{center}
\end{figure}
\begin{figure}
\hspace*{1cm}
\begin{center}
\epsfig{file=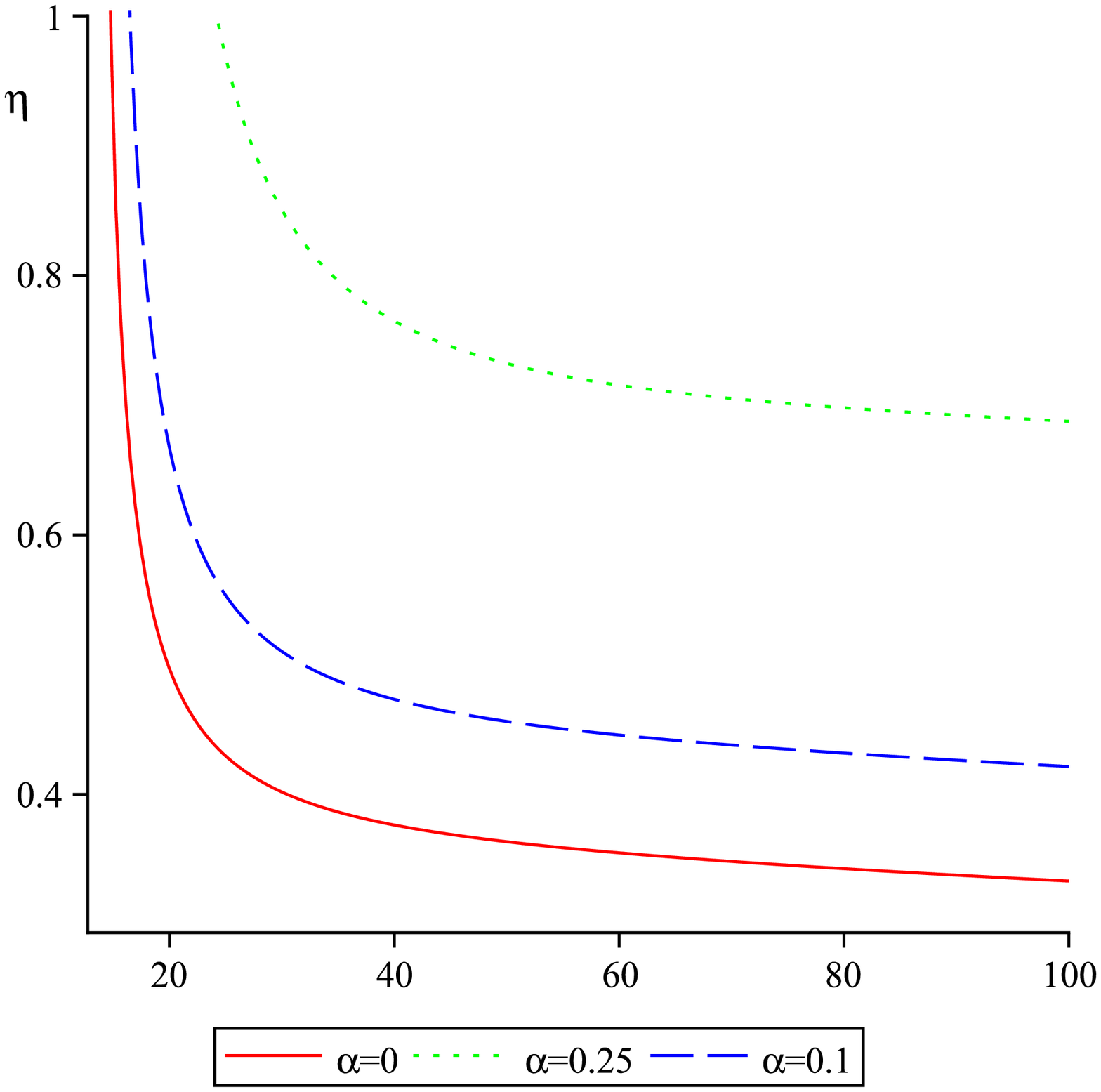,width=7cm}
\epsfig{file=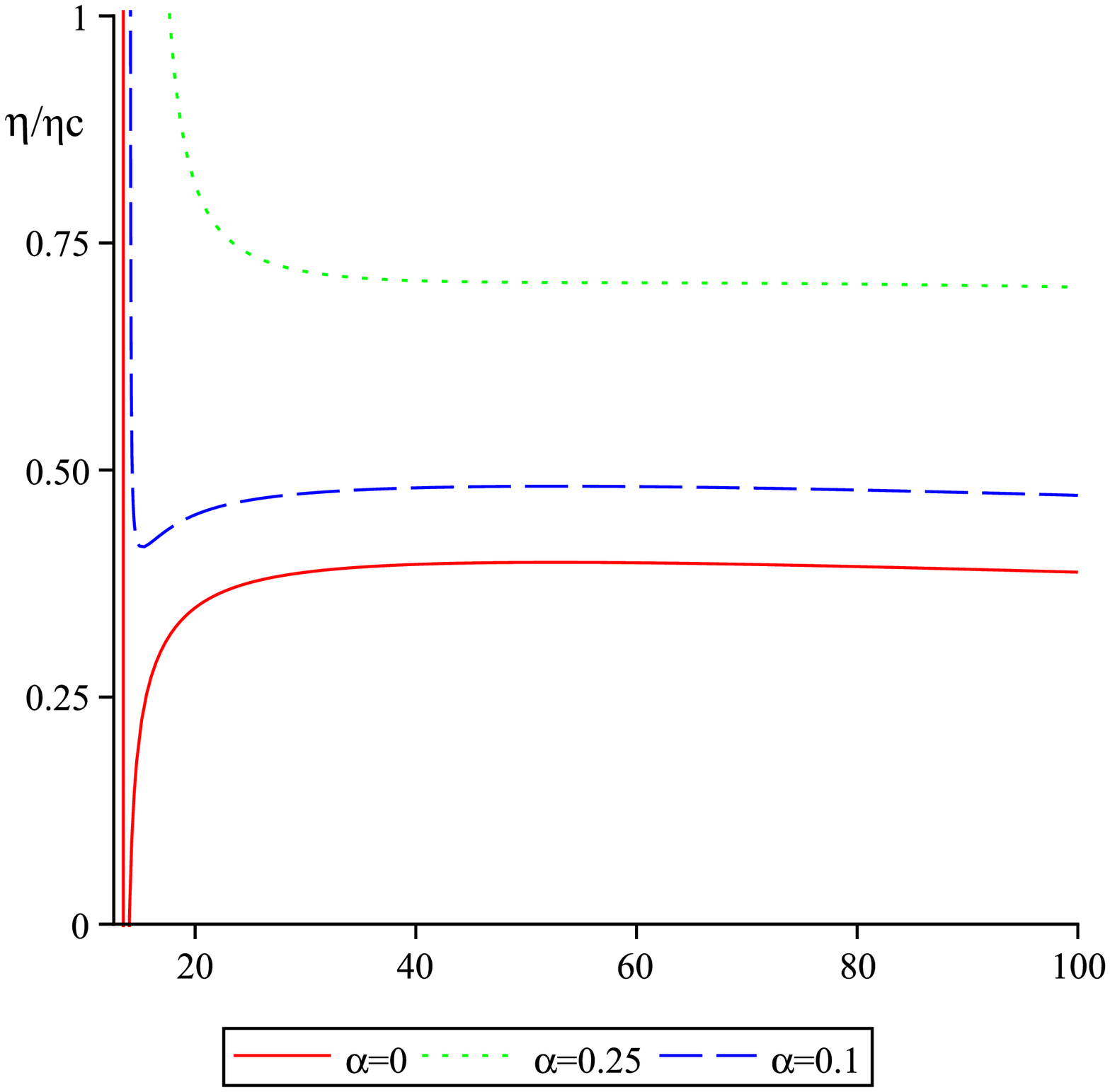,width=7cm}\caption{\small{Left plot: $\eta$
 with respect to $V_{R}$ for $J=0.1$, $Q=0.05$, $P_{T}=0.07$, $P_{B}=0.06$, $V_{L}=10$ and different values of $\alpha$ ; Right
 plot: $\frac{\eta}{\eta_{C}}$ with respect to $V_{R}$ for $J=0.1$, $Q=0.05$, $P_{T}=0.07$, $P_{B}=0.06$, $V_{L}=10$ and different values of $\alpha$
}}
\end{center}
\end{figure}
\begin{figure}
\hspace*{1cm}
\begin{center}
\epsfig{file=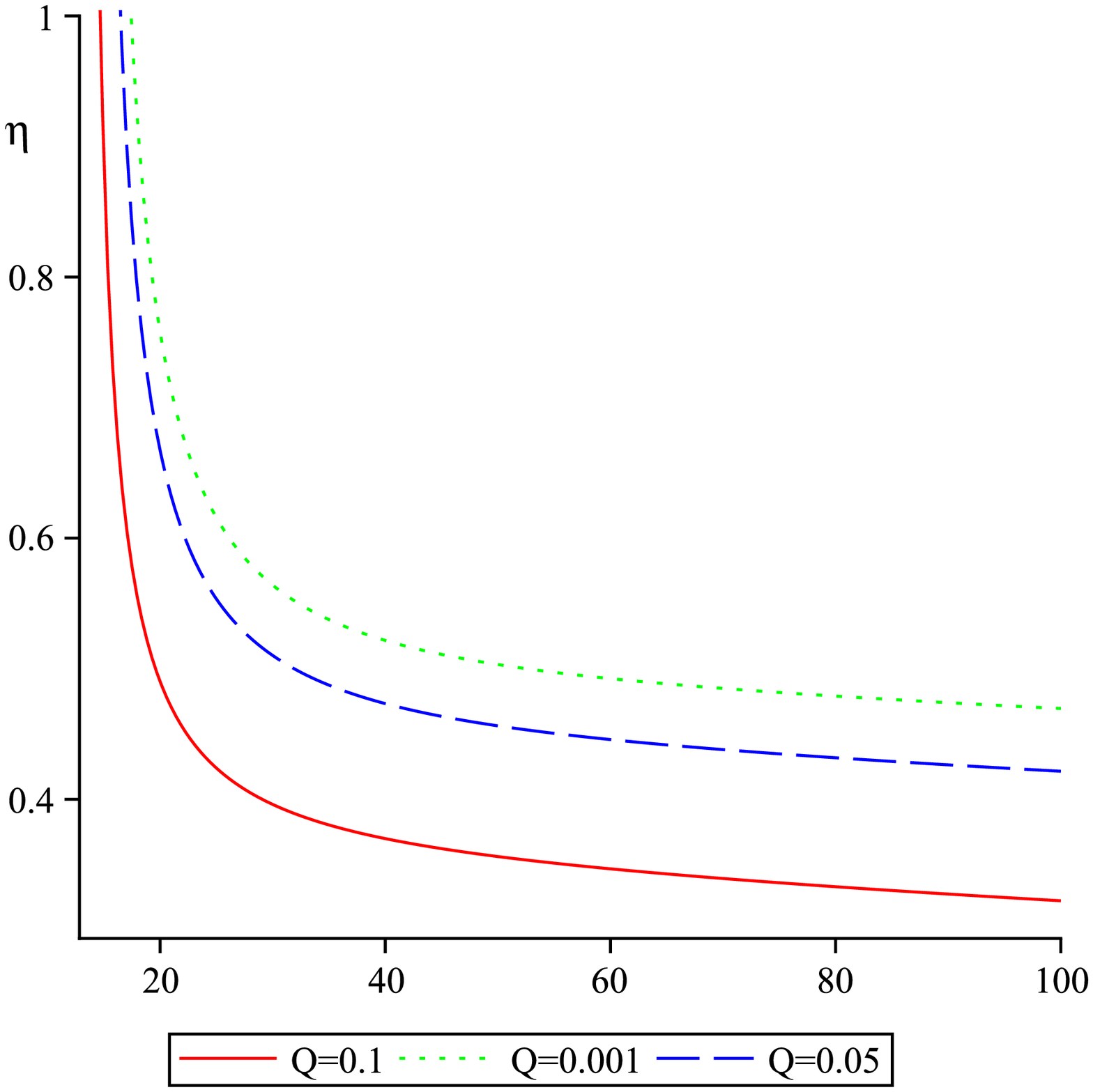,width=7cm}
\epsfig{file=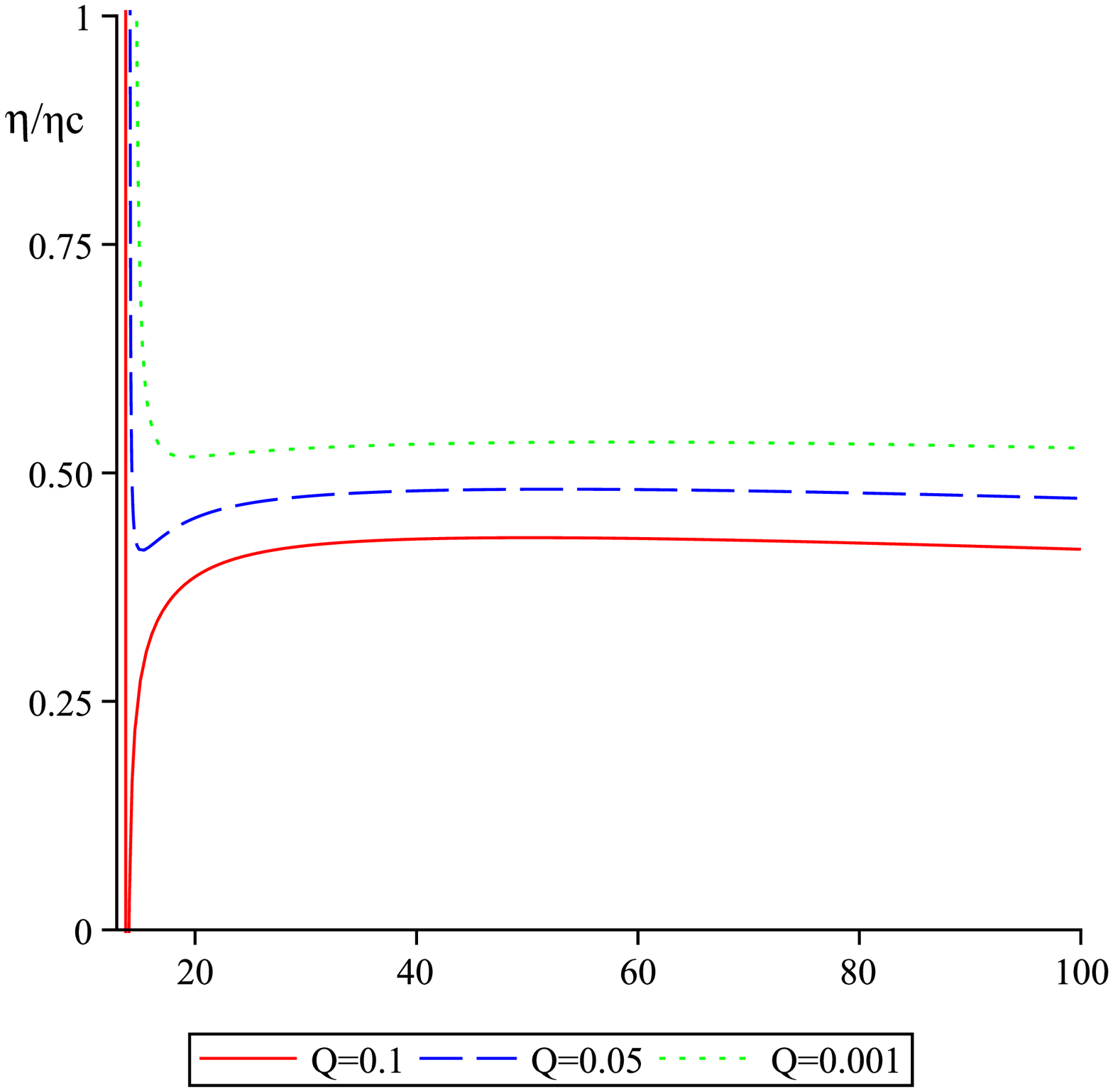,width=7cm}\caption{\small{Left plot: $\eta$
 with respect to $V_{R}$ for $\alpha=0.1$, $J=0.1$, $P_{T}=0.07$, $P_{B}=0.06$, $V_{L}=10$ and different values of $Q$ ; Right
 plot: $\frac{\eta}{\eta_{C}}$ with respect to $V_{R}$ for $\alpha=0.1$, $J=0.1$, $P_{T}=0.07$, $P_{B}=0.06$, $V_{L}=10$ and different values of $Q$
}}
\end{center}
\end{figure}

\begin{figure}
\hspace*{1cm}
\begin{center}
\epsfig{file=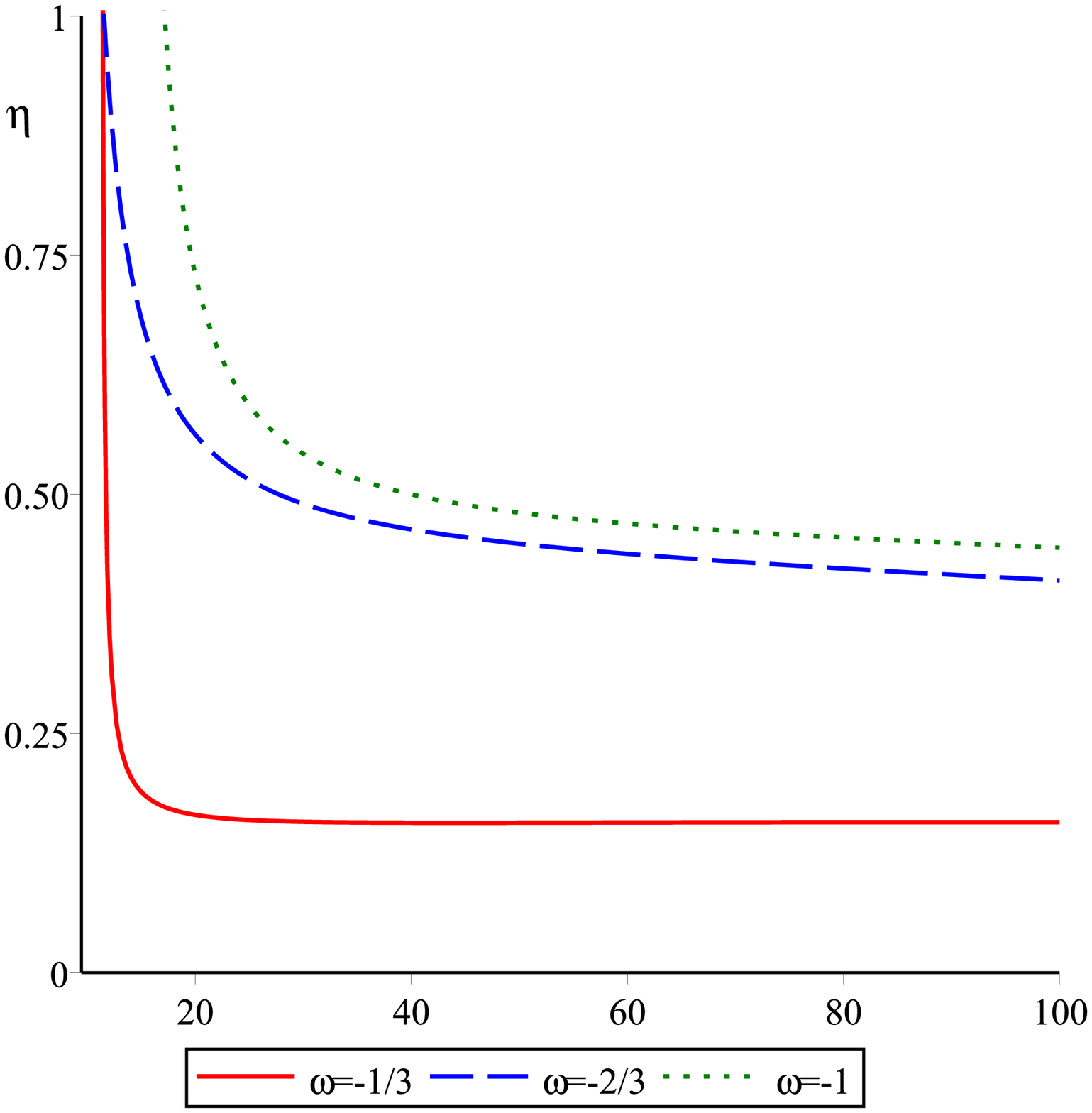,width=7cm}
\epsfig{file=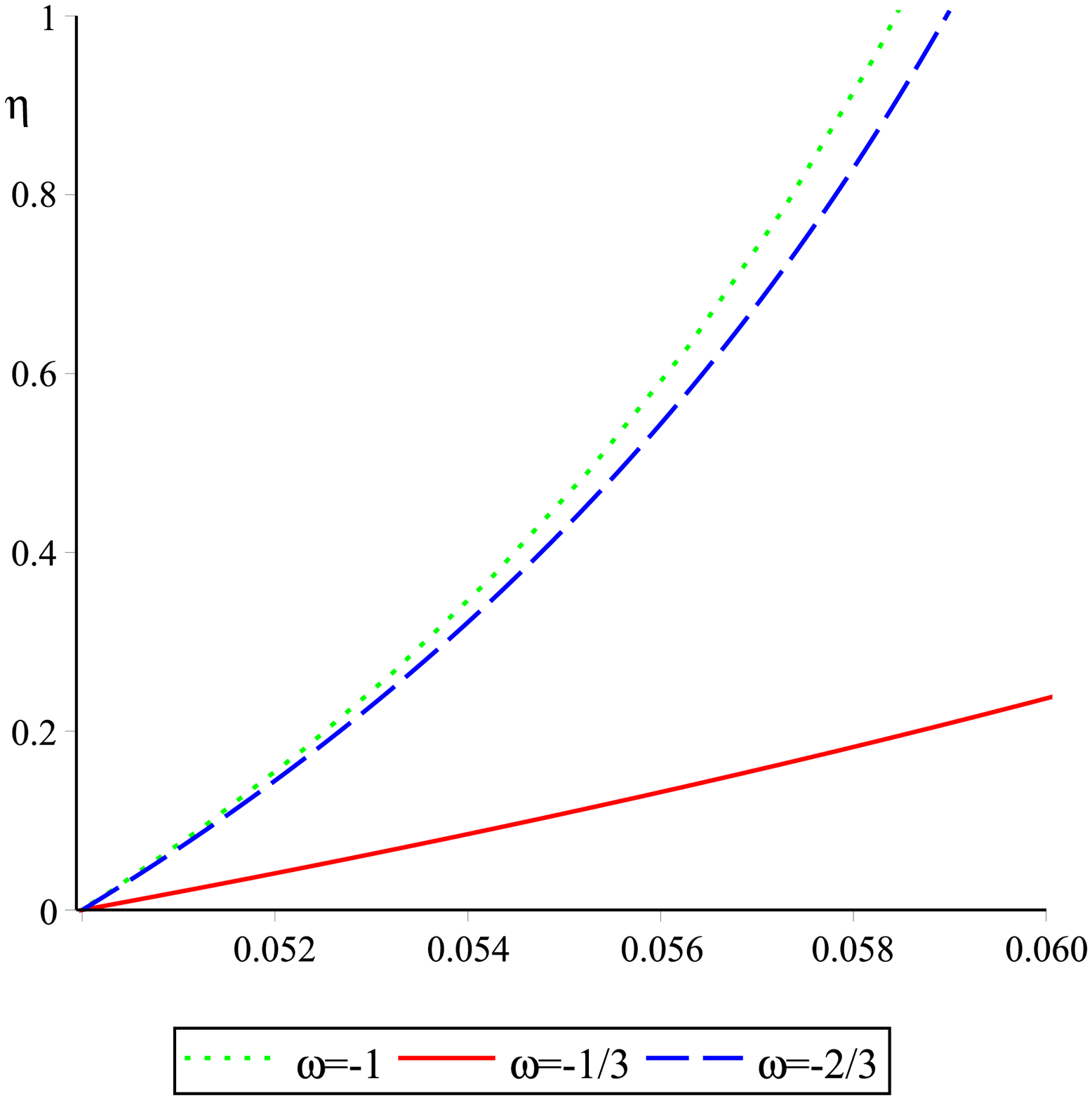,width=7cm}\caption{\small{Left plot: $\eta$
 with respect to $V_{R}$ for $\alpha=0.1$, $J=0.1$, $Q=0.05$, $P_{B}=0.07$, $P_{B}=0.06$, $V_{L}=10$ and different values of $\omega$ ; Right
 plot: $\eta$
 with respect to $P_{T}$ for $\alpha=0.1$, $Q=0.05$, $J=0.1$, $P_{B}=0.05$, $V_{R}=20$, $V_{L}=10$ and different values of $\omega$}}
\end{center}
\end{figure}
\section{The efficiency of heat engine of corresponding black hole}
In this section, we are going to investigate the effects of
quintessence field on the efficiency of Kerr-Newman AdS black hole.
We need to calculate heat capacity under constant pressure and
constant volume to obtain efficiency of black hole. For the static
holes, the thermodynamic volume and the entropy are not independent.
So, the heat capacity in constant volume ($C_{V}=T\frac{\partial
S}{\partial T}\large\mid_{V}$) is zero. It shows that the heat flow
occur along the isobars and one can obtain the efficiency by
straightforward approach as Ref [40]. But for rotation black holes
the thermodynamic volume and the entropy are independent and the
heat capacity is non-zero under constant volume. So, heat flow occur
along the isobars and isochores. In this case one cannot employ
usual methods for obtaining efficiency. Here we use the mentioned
approach in Ref. [56] and obtain the efficiency as,

\begin{equation}
\eta=\frac{W}{Q_{H}}=1-\frac{Q_{C}}{Q_{H}},
\end{equation}
where $Q_{H}$, $Q_{C}$ and $W$ are net input heat flow, net output
heat flow and net output work respectively. In the case of a
rectangular cycle for rotating black holes with $C_{V}\neq0$,
$Q_{H}$ and $Q_{C}$ are expressed by following expression,

\begin{eqnarray}
Q_{C}=M(V_{R},P_{T})-M(V_{L},P_{B})-\Delta PV_{R}\nonumber
\\&& \hspace{-79mm}
Q_{H}=M(V_{R},P_{T})-M(V_{L},P_{B})-\Delta PV_{L},
\end{eqnarray}
where $\Delta P=P_{T}-P_{B}$. By substituting (32) to (31), the
efficiency is determined as,
\begin{equation}
\eta=\frac{\Delta P \Delta V}{\Delta M_{T}+\Delta U_{L}},
\end{equation}
where $\Delta V=V_{R}-V_{L}$, $U=M-PV$ and

\begin{eqnarray}
\Delta M_{T}=M(V_{R},P_{T})-M(V_{L},P_{T})\nonumber
\\&& \hspace{-67mm}
\Delta U_{L}=U(V_{L},P_{T})-M(V_{L},P_{B}).
\end{eqnarray}
The efficiency of engine cannot be larger than the Carnot
efficiency. Because in this case the second law will be violated.
The efficiency of the Carnot engine is obtained by following,
\begin{equation}
\eta_{C}=1-\frac{T_{C}}{T_{H}}=1-\frac{T(P_{B},V_{L})}{T(P_{T},V_{R})}.
\end{equation}
Here important subject is that how the efficiency changes by varying
normalization factor $\alpha$, state parameter $\omega$, angular
momentum $J$ and charge of black hole. We have plotted efficiency
$\eta$ and  ratio $\frac{\eta}{\eta_{C}}$ for $\omega=-\frac{1}{3}$
in figures (4-6). Figure (4) shows differentiation of $\eta$ and
$\frac{\eta}{\eta_{C}}$ with respect to $V_{R}$ for different values
of $J$. As we see, $\eta$ increases by increasing $J$ but it
monotonously decrease with $V_{R}$. In other words, when the volume
difference between small black hole ($V_{L}$) and large black hole
($V_{R}$) is increased, the efficiency is declined until it reaches
to a constant value. Also, we see  when the angular momentum rises
the diagram of $\eta$ shifts to a larger $V_{R}$. It means that  by
increasing $J$ the volume of large black hole increases therefore
$\Delta V$ (the volume difference between small black hole and large
black hole) will be larger. The right figure of Fig 4. is similar to
the left one. It shows that $\frac{\eta}{\eta_{C}}$ monotonously
decreases by increasing $V_{R}$ which means that the difference
between the actual efficiency ($\eta$) and ($\eta_{C}$) will be more
for larger $\Delta V$. Also, we notice that
$\frac{\eta}{\eta_{C}}<1$ holds for $J<0.5$ so the second law of the
thermodynamics would violate for $J>0.5$.\\
The Fig. 5 has been plotted for three different normalization factor
$\alpha$. As we see, the curves are very similar to the Fig.4, then
we can use same discussions in this case. Also we notice that the
length of the defined cycle significantly increases for $1<\alpha<2$
i.e. $\Delta V$ highly rises in this range. The right figure in
Fig.5 shows that the second law is satisfied for $\alpha<2$. We have
plotted figures of $\eta$ and $\frac{\eta}{\eta_{C}}$ for three
values of charge in Fig. 6. The discussions are somewhat similar to
figures of (4) and (5) but the difference is that $\eta$ increases
by decreasing $Q$ and the second law holds for any value of charge.
We continue our analysis for $\omega=-\frac{2}{3}$. The results are
almost identical to the previous one. But there is a significant
difference in this case. As we notice that in Fig. 7
$\frac{\eta}{\eta_{C}}$ reduces by decreasing $J$. But for very
small values of $J$ the difference between $\eta$ and $\eta_{C}$ is
such that $\frac{\eta}{\eta_{C}}$ first decreases afterward
increases and then it reduces and eventually it reaches to a
constant value. One can use similar explanation for right figure of
Fig. 9. Also we investigate behavior of $\eta$ and
$\frac{\eta}{\eta_{C}}$ under changes of $J$, $\alpha$ and $Q$ for
$\omega=-1$ in figures (10-12). Our analysis is same as case of
$\omega=-\frac{2}{3}$. Finally we investigate our the $\eta$ under
change of pressure $P_{T}$. The figure (13) shows that the $\eta$
monotonously increase as $P_{T}$ rises. And by reducing $\omega$
from $-\frac{1}{3}$ to $-1$ the efficiency increases. Finally, by
comparing these figures we notice when the state parameter $\omega$
increases from $-1$ to $-\frac{1}{3}$ range of $J$ and $\alpha$
(which can satisfy the second law) rises.
\section{conclusion}
In this paper, we studied $P-V$ criticality of Kerr-Newman $AdS$
black hole with a quintessence field. We calculated critical
quantities and saw that by reducing the state parameter $\omega$
from $-\frac{1}{3}$ to $-1$ the critical pressure and temperature
increases. Also we obtained universal ratio and noticed
 for the state parameter
$\omega= -\frac{1}{3}$, the obtained ratio is quite same as
Kerr-Newman $AdS$ black hole without dark energy parameter. Then, we
considered corresponding black hole and investigated the influence
of quintessence field $\alpha$, $\omega$ and $J$ on the efficiency
$\eta$. We found that the $\eta$ increases by increasing $J$ and
$\alpha$ and reducing charge $Q$ of black hole. Also we saw that the
$\eta$ monotonously decrease by rising the volume of large black
hole ($V_{R}$) i.e. when volume difference between small black hole
and large black hole ($\Delta V$) grows then the $\eta$ highly
reduces. We found that when $J$ and $\alpha$ increases the length of
defined cycle increases, in other words $\Delta V$ increases. By
studying ratio the $\eta$ and the highest efficiency (Carnot
efficiency $\eta_{C}$) we noticed that the second law of the
thermodynamics will be violated for special values of $J$ and
$\alpha$ but it holds for any value of $Q$. It means that constraint
of $\frac{\eta}{\eta_{C}}<1$ is not satisfied for any value of $J$
and $\alpha$. And also by increasing $\omega$ from $-1$ to
$-\frac{1}{3}$ the range of $J$ and $\alpha$ which satisfy the
second law increases. We showed that when $\omega$ increases from
$-1$ to $-\frac{1}{3}$ then the efficiency decreases. Finally, we
investigated behavior the $\eta$ under changes of pressure. We saw
that by increasing pressure the $\eta$ monotonously increases and
the second law is violated by increasing pressure difference between
$P_{B}$ and $P_{T}$ in other words the second law holds for very
small $\Delta P$.

\end{document}